\documentclass[
final
]{elsarticle}
\usepackage[
reviewcopy
]{adndt}

\usepackage{longtable}

\usepackage[utf8]{inputenc}
\usepackage[T1]{fontenc}
\usepackage{microtype}

\usepackage{xcolor}
\usepackage[colorlinks=true,linkcolor=blue,urlcolor=blue,citecolor=blue,bookmarks=true,bookmarksopenlevel=2,breaklinks=true]{hyperref}

\usepackage{mathptmx}

\usepackage{amsmath}
\usepackage{amssymb}
\usepackage{array,booktabs,tabularx}

\biboptions{square,sort&compress}
\bibpunct[]{[}{]}{,}{n}{}{;}
\usepackage{xspace}
\newcommand{\grasp}{\textsc{Grasp2018}\xspace}
\newcommand{\fac}{\textsc{Fac}\xspace}

\newcommand{\cmm}{cm\textsuperscript{-1}\xspace}

\usepackage{mhchem}
\usepackage{siunitx}
\usepackage[normalem]{ulem}
\usepackage{soul}
\soulregister\cite7
\soulregister\ref7
\soulregister\pageref7

\usepackage{tikz}
\usepackage{pgfplots}
\pgfplotsset{compat=1.8}
\usepgfplotslibrary{dateplot}
\usepackage{sansmath}

\begin{document}

\begin{frontmatter}

\journal{Atomic Data and Nuclear Data Tables}


\title{Energies of \ce{I^8+} through \ce{I^12+} low-lying levels}


  \author{Karol Kozio{\l}\corref{cor}}
  \ead{E-mail: karol.koziol@ncbj.gov.pl}

  \cortext[cor]{Corresponding author.}

  \author{Jacek Rzadkiewicz}

  \address{Narodowe Centrum Bada\'{n} J\k{a}drowych (NCBJ), Andrzeja So{\l}tana 7, 05-400 Otwock-\'{S}wierk, Poland}

\date{2024} 

\begin{abstract}   
High-accuracy Multi-Configuration Dirac--Hartree--Fock with Configuration Interaction calculations of level energies and transition rates have been carried out for iodine \ce{I^8+} through \ce{I^12+} ions related to the \ce{[Kr]{4d}^n} ($n$ = 5--9) configurations. 
For \ce{I^10+} through \ce{I^12+} ions the present data fill up the lack of such data in the literature. 
\end{abstract}

\end{frontmatter}

\clearpage

\tableofcontents
\listofDtables
\listofDfigures
\vskip5pc


\section{Introduction}

The spectroscopic properties of high-charged ions (HCIs) with the open $d$ shell are widely utilized in fields such as astronomy and plasma physics. The presence of open $d$ shells leads to the existence of numerous atomic levels and radiative transitions. 
Recent study suggested that HCIs with $\text{d}^{(4-6)}$ configurations might host suitable forbidden transitions for optical-clock frequency measurements \cite{Lyu2023}. 
The low-lying states related to the $(n = 4 - 5)\text{d}^6$ and $(n = 4 - 5)\text{d}^8$ configurations has been analyzed by Yu and Sahoo \cite{Yu2024} for a series of ions across Mo-like, W-like, Ru-like, and Os-like isoelectronic sequences in order to identify transitions that can be suitable for making single-ion-based optical clocks, however iodine ions have been not considered in that study. 
Indeed there is a lack of data for energy levels of iodine ions with intermediate charge in the National Institute of Standards and Technology Atomic Spectra Database (NIST ASD) \cite{NIST_ASD}. 
The spectrum of I IX has been analyzed by Joshi et al. \cite{Joshi1980} and then by Churilov et al. \cite{Churilov1998}. The spectrum of I X has been analyzed by Gayasov et al. \cite{Gayasov1997} and then also by Churilov et al. \cite{Churilov1998}. 
Ivanova \cite{Ivanova2009} provided calculations for the fine structure energy splitting \ce{[Kr] {4d}^9 ^2D5/2-^2D3/2} for Rh-like ions with $52 \le Z \le 86$, including iodine ion. 
Unfortunately there is no detailed analysis of the I XI, I XII, and I XIII spectra in the literature. 
Here we present the high-accuracy calculations of, employing the Multi-Configuration Dirac--Hartree--Fock (MCDHF) method with Configuration Interaction (CI), of energy levels and transition rates between them for \ce{[Kr]{4d}^n} ($n$ = 5--9) configurations for I$^{8+}$ through I$^{12+}$ ions.

\section{Theoretical background}

The calculations of the radiative transition energies and rates have been carried out by means of the \textsc{Grasp2018} \cite{FroeseFischer2018} code.
The \textsc{Grasp2018} code is based on the MCDHF method. 
The methodology of MCDHF calculations performed in the present study is similar to that published earlier in many papers (see, e.g., \cite{Dyall1989,Grant2007}).
The effective Hamiltonian for an $N$-electron system is expressed by
\begin{equation}
H = \sum_{i=1}^{N} h_{D}(i) + \sum_{j>i=1}^{N} C_{ij},
\end{equation}
where $h_D(i)$ is the Dirac operator for the $i$th electron and the terms $C_{ij}$ account for the electron--electron interactions.
In general, the latter is a sum of the Coulomb interaction operator and the transverse Breit operator.
An atomic state function (ASF) with total angular momentum $J$ and parity $p$ is assumed in the form
\begin{equation}
\Psi_{s} (J^{p} ) = \sum_{m} c_{m} (s) \Phi ( \gamma_{m} J^{p} ),
\end{equation}
where $\Phi ( \gamma_{m} J^{p} )$ are the configuration state functions (CSFs), $c_{m} (s)$ are the configuration mixing coefficients for state $s$, and $\gamma_{m}$ represents all information required to define a certain CSF uniquely.
The CSFs are linear combinations of $N$-electron Slater determinants which are antisymmetrized products of 4-component Dirac orbital spinors. 
In the \textsc{Grasp2018} code, the Breit interaction contribution to the energy (calculated in low-frequency limit) is added perturbatively, after the radial part of wavefunction has been optimized. 
Also two types of quantum electrodynamics (QED) corrections: the self-energy (as the screened hydrogenic approximation \cite{McKenzie1980} of the data of Mohr and co-workers \cite{Mohr1992a}) and the vacuum polarization (as the potential of Fullerton and Rinker \cite{Fullerton1976}) have been included. 
The accuracy of the wavefunction depends on the CSFs included in its expansion \cite{FroeseFischer2014,FroeseFischer2016}.
The accuracy can be improved by extending the CSF set by including the CSFs originating from excitations from orbitals occupied in the reference CSFs to unfilled orbitals of the active orbital set (i.e., CSFs for virtual excited states).
This approach is called Configuration Interaction (CI). 
The CI method makes it possible to include the major part of the electron correlation contribution to the energy of the atomic levels.
In the CI approach, it is very important to choose an appropriate basis of CSFs for the virtual excited states.
It can be done by systematically building CSF sequences by extending the Active Space (AS) of orbitals and concurrently monitoring the convergence of the self-consistent calculations. 

\begin{table*}[!htb]
\caption{\label{tab:cfs-no}Active spaces of CSFs used in calculations.}
\centering
\begin{tabular}{ll rrrrr}
\toprule
& & \multicolumn{5}{c}{Number of CSFs}\\
\cmidrule{3-7}
Active space & Virtual orbitals & I$^{8+}$ & I$^{9+}$ & I$^{10+}$ & I$^{11+}$ & I$^{12+}$\\
\midrule	
AS0 & & 2 & 9 & 19 & 34 & 37 \\
AS1 & 4f + $n=5$, $l=0{-}5$ & 13329 & 80991 & 199886 & 302780 & 316733 \\
AS2 & 4f + $n=5{-}6$, $l=0{-}5$ & 42778 & 258780 & 636743 & 958366 & 993778 \\
AS3 & 4f + $n=5{-}7$, $l=0{-}5$ & 88907 & 536698 & 1318954 & 1980512 & 2047144 \\
\bottomrule
\end{tabular}
\end{table*}

The active spaces used in calculations for particular iodine ions are presented in Table~\ref{tab:cfs-no}. 
The AS0 is a multi-reference (MR) CSFs set. The MR configurations are \ce{[Kr]{4d}^n} where $n$ = 9, 8, 7, 6, 5 for I$^{8+}$, I$^{9+}$, I$^{10+}$, I$^{11+}$, I$^{12+}$ ions respectively. 
We have considered all possible single (S) and double (D) substitutions from the 4s, 4p, 4d occupied subshells to the active space of virtual orbitals. 
In this case, the inactive core contains $n = 1, 2, 3$ shells. 
In our calculations we used the active spaces of virtual orbitals with $n$ up to $n=7$ and $l$ up to $l=5$. 
From Table~\ref{tab:cfs-no} one can conclude that: (i) the number of CSFs increases rapidly when active space grows, reaching $\sim2\times10^6$ CSFs for AS3 for I$^{11+}$ (4d$^6$) and I$^{12+}$ (4d$^5$) ions, and (ii) the size of the expansions increases with the size of the reference set.

\section{Results}

\subsection{Probing the convergence of MCDHF-CI calculations}
\label{sec:corr}

\begin{figure}[!htb]
\centering
\includegraphics[width=0.5\linewidth]{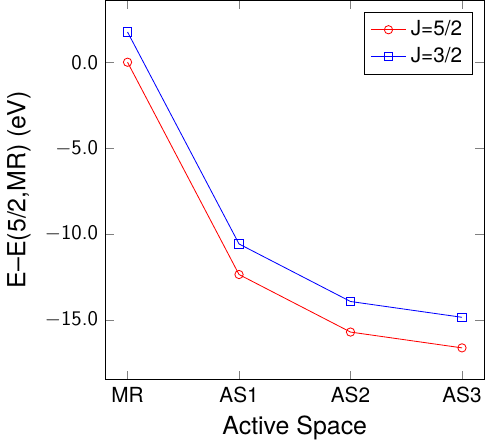}
\caption{Level energy convergence in CI calculation for I$^{8+}$ -- shifted absolute energies.\label{fig:I8+-conv}}
\end{figure}

\begin{figure}[!htb]
\centering
\includegraphics[width=0.5\linewidth]{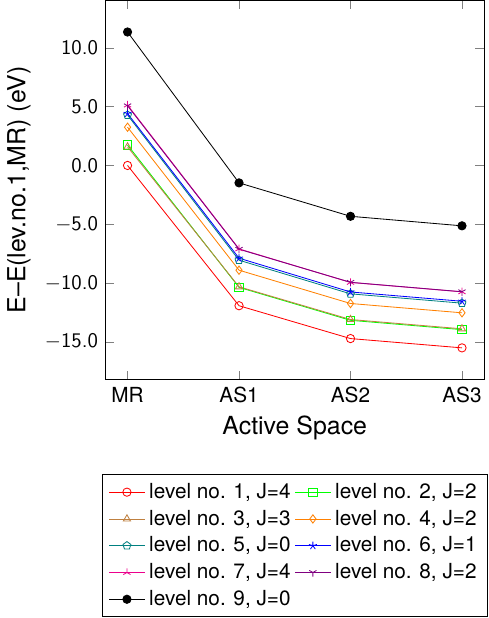}
\caption{Level energy convergence in CI calculation for I$^{9+}$ -- shifted absolute energies.\label{fig:I9+-conv}}
\end{figure}

\begin{figure}[!htb]
\centering
\includegraphics[width=0.5\linewidth]{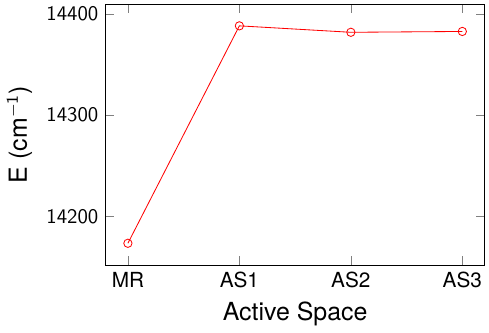}
\caption{Level energy convergence in CI calculation for $J=3/2$ level of I$^{8+}$ -- relative energy.\label{fig:I8+-conv-rel}}
\end{figure}

\begin{figure}[!htb]
\centering
\includegraphics[width=0.5\linewidth]{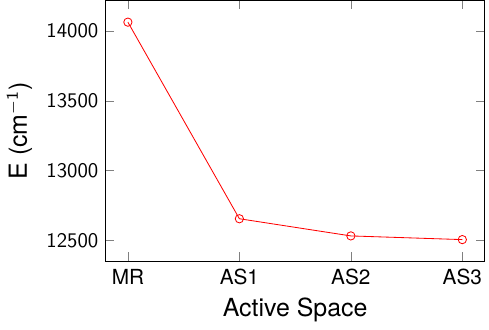}
\caption{Level energy convergence in CI calculation for the level no. 2 of I$^{9+}$ -- relative energy.\label{fig:I9+-conv-rel_l2}}
\end{figure}

\begin{figure}[!htb]
\centering
\includegraphics[width=0.5\linewidth]{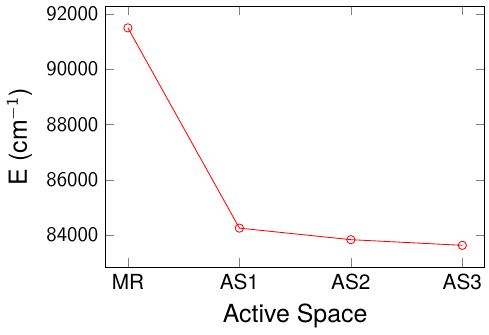}
\caption{Level energy convergence in CI calculation for the level no. 9 of I$^{9+}$ -- relative energy.\label{fig:I9+-conv-rel_l9}}
\end{figure}

Figures~\ref{fig:I8+-conv} and~\ref{fig:I9+-conv} present the convergence in MCDHF-CI calculations for levels of I$^{8+}$ and I$^{9+}$ ions. The energy values presented therein are the level energies relative to the absolute energy of the ground state calculated at the MR stage, i.e $\tilde{E}(\text{state\#}x,\text{AS}y)=E(\text{state\#}x,\text{AS}y)-E(\text{state\#}1,\text{MR})$. Looking at Figures~\ref{fig:I8+-conv} and~\ref{fig:I9+-conv} one can claim that the AS3 stage is adequate to achieve level energy convergence. 
The convergence of MCDHF-CI calculations near to the AS3 stage is clearer visible when considering energies relative to the groundstate, i.e. $\bar{E}(\text{state\#}x,\text{AS}y)=E(\text{state\#}x,\text{AS}y)-E(\text{state\#}1,\text{AS}y)$ -- see Figures~\ref{fig:I8+-conv-rel}, \ref{fig:I9+-conv-rel_l2}, and \ref{fig:I9+-conv-rel_l9} for the cases of selected states of I$^{8+}$ and I$^{9+}$ ions. 

\subsection{Estimating theoretical uncertainties}
\label{sec:unc_theo}

The theoretical uncertainties of level energies are related to convergence with the size of a basis set and estimated as absolute value of difference between energies calculated within AS2 and AS3, i.e. $\delta E = |E^\text{AS3}-E^\text{AS2}|$. 
Such an estimation gives 0.8~\cmm (0.005\%) uncertainty for the energy of \ce{^2D5/2-^2D3/2} splitting in I$^{8+}$, up to 0.3\% uncertainty for theoretical level energies of I$^{9+}$ and I$^{10+}$ ions, and up to 0.5\% uncertainty for theoretical level energies of I$^{11+}$ and I$^{12+}$ ions. 
The theoretical uncertainties of transition rates were estimated according to the procedure described by Kramida \cite{Kramida2013}. 
Among all M1 transitions for given ion, the subset of high-intensity transitions (with the intensities higher than 10\% of the strongest transition intensity) has been selected for estimating transition rate uncertainty, i.e. $\delta A = |A^\text{AS3}-A^\text{AS2}|$. The obtained value of uncertainty has been extended to all studied transitions. 
Such an estimation gives $\approx$0.02\% uncertainty for transition rate of \ce{^2D5/2-^2D3/2} transition in I$^{8+}$ and $\approx$0.3\% uncertainty for all transition rates of I$^{9+}$, I$^{10+}$, I$^{11+}$, I$^{12+}$ ions. 

\begin{table}[!htb]
\caption{High-intensity M1 transitions selected for estimating transition rate uncertainty, by comparing the \textsc{Grasp2018} and the \textsc{Fac} calculated numbers.\label{tab:comp-grasp-fac}}
\centering
\begin{tabular}{@{} ll SSS @{}}
\toprule
Ion & Transition & \multicolumn{2}{c}{Transition rate (s$^{-1}$)} & {$|\Delta|$ (\%)} \\
\cmidrule{3-4}
&& {\textsc{Grasp}} & {\textsc{Fac}} &\\
\midrule	
I$^{8+}$ & \ce{^2D5/2-^2D3/2} & 48.09 & 45.29 & 5.8 \\
I$^{9+}$ & \ce{^1S0-^3P1} & 637.21 & 655.31 & 2.8 \\
I$^{9+}$ & \ce{^1D2-^3F3} & 105.33 & 98.76 & 6.2 \\
I$^{9+}$ & \ce{^3F3-^3F4} & 56.32 & 51.46 & 8.6 \\
I$^{10+}$ & \ce{^2D5/2-^4F7/2} & 222.86 & 223.77 & 0.4 \\
I$^{10+}$ & \ce{^2D3/2-^2D5/2} & 177.43 & 162.52 & 8.4 \\
I$^{10+}$ & \ce{^2D5/2-^4P5/2} & 162.10 & 164.99 & 1.8 \\
I$^{11+}$ & \ce{^1S0-^3P1} & 765.77 & 764.77 & 0.1 \\
I$^{11+}$ & \ce{^3P0-^3D1} & 313.55 & 306.29 & 2.3 \\
I$^{11+}$ & \ce{^3P0-^3D1} & 299.43 & 250.88 & 16.2 \\
I$^{12+}$ & \ce{^4F5/2-^4G5/2} & 258.74 & 295.83 & 14.3 \\
I$^{12+}$ & \ce{^2P1/2-^2S1/2} & 212.57 & 226.43 & 6.5 \\
I$^{12+}$ & \ce{^2F5/2-^4F5/2} & 177.18 & 198.28 & 11.9 \\
\bottomrule
\end{tabular}
\end{table}

Where there are no reference values to which the calculated numbers can be compared to, the comparison between numbers calculated by two difference methodologies can be used to estimated the theoretical differences for the transition rates. Table~\ref{tab:comp-grasp-fac} presents the values for three strongest M1 transition rates for each I ion (except for I$^{8+}$, for which only one transition is analyzed), calculated by means of the \textsc{Grasp2018}, based on the MCDHF method, and by means of the \textsc{Fac} code \cite{Gu2008}, based on the multiconfigurational Dirac--Hartree--Fock--Slater (MCDHFS) method. 
The factor $\left|\Delta\right|=\left|\left(A^\text{GRASP}-A^\text{FAC}\right)/A^\text{GRASP}\right|$ evaluates the difference between \grasp and \fac results. 
One can see from Table~\ref{tab:comp-grasp-fac} that the transition rates for I$^{8+}$, I$^{9+}$, and I$^{10+}$ ions have B accuracy class (uncertainties $\le 10\%$), according to accuracy categorization introduced in the NIST ASD, while the transition rates for I$^{11+}$ and I$^{12+}$ ions have C+ accuracy class (uncertainties $\le 18\%$).


\subsection{Comparing to experiments}
\label{sec:unc}

\begin{table}[!htb]
\caption{\label{tab:comp-exp-i8+}Comparison of present MCDHF-CI calculation results for \ce{[Kr] {4d}^9 ^2D5/2-^2D3/2} energy splitting in I$^{8+}$ ion to the experimental data found in the literature. Relative difference between theoretical and experimental data, given as $\Delta = (\text{Theo.}-\text{Exp.})/\text{Exp.}$ is also reported.}
\centering
\begin{tabular}{@{} SS l S @{}}
\toprule
{Theoretical (\cmm)} & {Experimental (\cmm)} & Ref. & {$\Delta$ (\%)} \\
\midrule	
14382.51 & 14362 &\cite{Joshi1980} & 0.1\\
& 14392.77 &\cite{Kimura2023} & -0.1\\
\bottomrule
\end{tabular}
\end{table}

\begin{table}[!htb]
\caption{\label{tab:comp-exp-i9+}Present MCDHF-CI calculated energy levels for 4d$^8$ valence electronic configuration of I$^{9+}$ ion compared to the ones deduced in Ref.~\cite{Churilov1998} from experimental spectra. Relative difference between theoretical and experimental data, given as $\Delta = (\text{Theo.}-\text{Exp.})/\text{Exp.}$ is also reported.}
\centering
\begin{tabular}{@{} l SSS @{}}
\toprule
$J$ & {Theoretical (\cmm)} & {Experimental (\cmm), Ref.~\cite{Churilov1998}} & {$\Delta$ (\%)} \\
\midrule
4 & 0.0 & 0 & {--}\\
2 & 12505.90 & 12230 & 2.3 \\
3 & 13051.94 & 13043 & 0.1 \\
2 & 24030.07 & 23854 & 0.7 \\
0 & 30638.93 & 29860 & 2.6 \\
1 & 31967.99 & 31440 & 1.7 \\
4 & 38430.16 & 37503 & 2.5 \\
2 & 38488.11 & 38198 & 0.8 \\
0 & 83632.17 & 81670 & 2.4 \\
\bottomrule
\end{tabular}
\end{table}

Present MCDHF-CI calculated energy levels for 4d$^9$ valence electronic configuration of I$^{8+}$ ion and 4d$^8$ valence electronic configuration of I$^{9+}$ ion compared to the ones deduced from experimental spectra are presented in Tables~\ref{tab:comp-exp-i8+} and~\ref{tab:comp-exp-i9+}. As one can see the theory--experiment difference does not exceed 2.6\%. Basing on this finding and on theoretical uncertainties presented in Section~\ref{sec:unc_theo} we roughly estimate the uncertainties for energy levels for I$^{9+}$ through I$^{12+}$ to be 3\% (A accuracy class).

%

%

\clearpage

\TableExplanation

\section*{Tables 1-5.}

\begin{tabular}{@{}p{0.5in}p{6in}@{}}
$J$		& Total momentum of the state\\
$p$		& Parity of the state\\
$E$		& Energy of atomic level relative to the groundstate (\cmm/eV)\\
\end{tabular}

\section*{Tables 6-10.}

\begin{tabular}{@{}p{0.5in}p{6in}@{}}
$\lambda$ & Transition wavelength (nm)\\
$A$ & Transition rate for magnetic-type transitions (s$^{-1}$)\\
\end{tabular}

\medskip

\noindent
Only transitions with $A>10~\text{s}^{-1}$ are listed.

\datatables 

\setlength{\LTcapwidth}{\linewidth}
\setlength{\LTleft}{0pt}
\setlength{\LTright}{0pt} 
\setlength{\tabcolsep}{0.5\tabcolsep}
\renewcommand{\arraystretch}{1.0}

\begin{longtable}{@{}@{\extracolsep{\fill}} lll p{0.65\linewidth} rr@{}}
\caption{I$^{8+}$ levels, 4d$^9$ valence electronic configuration (Rh-like ion).}\label{tab:ene-i8+}
No. & $J$ & $p$ & $LS$-composition &  $E$ (cm$^{-1}$) & $E$ (eV) \\ 
\midrule
\endfirsthead
\toprule
No. & $J$ & $p$ & $LS$-composition &  $E$ (cm$^{-1}$) & $E$ (eV) \\ 
\midrule
\endhead
\midrule
\endfoot
\bottomrule
\endlastfoot
1   & 5/2   & +   & 0.96~$3d^{10}(^{1}_{0}S)\,4s^{2}\,4p^{6}\,4d^{9}~^{2}D$                                    & 0            & 0.0000   \\ 
2   & 3/2   & +   & 0.96~$3d^{10}(^{1}_{0}S)\,4s^{2}\,4p^{6}\,4d^{9}~^{2}D$                                    & 14383        & 1.7832   \\
\end{longtable}

\clearpage

\begin{longtable}{@{}@{\extracolsep{\fill}} lll p{0.65\linewidth} rr@{}}
\caption{I$^{9+}$ levels, 4d$^8$ valence electronic configuration (Ru-like ion).}\label{tab:ene-i9+}
No. & $J$ & $p$ & $LS$-composition &  $E$ (cm$^{-1}$) & $E$ (eV) \\ 
\midrule
\endfirsthead
\toprule
No. & $J$ & $p$ & $LS$-composition &  $E$ (cm$^{-1}$) & $E$ (eV) \\ 
\midrule
\endhead
\midrule
\endfoot
\bottomrule
\endlastfoot
1   & 4     & +   & 0.94~$3d^{10}(^{1}_{0}S)\,4s^{2}\,4p^{6}\,4d^{8}(^{3}_{2}F)~^{3}F$ + 0.02~$3d^{10}(^{1}_{0}S)\,4s^{2}\,4p^{6}\,4d^{8}(^{1}_{2}G)~^{1}G$ & 0            & 0.0000   \\ 
2   & 2     & +   & 0.40~$3d^{10}(^{1}_{0}S)\,4s^{2}\,4p^{6}\,4d^{8}(^{1}_{2}D)~^{1}D$ + 0.38~$3d^{10}(^{1}_{0}S)\,4s^{2}\,4p^{6}\,4d^{8}(^{3}_{2}F)~^{3}F$ + 0.18~$3d^{10}(^{1}_{0}S)\,4s^{2}\,4p^{6}\,4d^{8}(^{3}_{2}P)~^{3}P$ & 12506        & 1.5505   \\ 
3   & 3     & +   & 0.96~$3d^{10}(^{1}_{0}S)\,4s^{2}\,4p^{6}\,4d^{8}(^{3}_{2}F)~^{3}F$                         & 13052        & 1.6182   \\ 
4   & 2     & +   & 0.53~$3d^{10}(^{1}_{0}S)\,4s^{2}\,4p^{6}\,4d^{8}(^{3}_{2}P)~^{3}P$ + 0.41~$3d^{10}(^{1}_{0}S)\,4s^{2}\,4p^{6}\,4d^{8}(^{3}_{2}F)~^{3}F$ & 24030        & 2.9793   \\ 
5   & 0     & +   & 0.88~$3d^{10}(^{1}_{0}S)\,4s^{2}\,4p^{6}\,4d^{8}(^{3}_{2}P)~^{3}P$ + 0.08~$3d^{10}(^{1}_{0}S)\,4s^{2}\,4p^{6}\,4d^{8}(^{1}_{0}S)~^{1}S$ & 30639        & 3.7987   \\ 
6   & 1     & +   & 0.96~$3d^{10}(^{1}_{0}S)\,4s^{2}\,4p^{6}\,4d^{8}(^{3}_{2}P)~^{3}P$                         & 31968        & 3.9635   \\ 
7   & 4     & +   & 0.94~$3d^{10}(^{1}_{0}S)\,4s^{2}\,4p^{6}\,4d^{8}(^{1}_{2}G)~^{1}G$ + 0.02~$3d^{10}(^{1}_{0}S)\,4s^{2}\,4p^{6}\,4d^{8}(^{3}_{2}F)~^{3}F$ & 38430        & 4.7647   \\ 
8   & 2     & +   & 0.54~$3d^{10}(^{1}_{0}S)\,4s^{2}\,4p^{6}\,4d^{8}(^{1}_{2}D)~^{1}D$ + 0.25~$3d^{10}(^{1}_{0}S)\,4s^{2}\,4p^{6}\,4d^{8}(^{3}_{2}P)~^{3}P$ + 0.17~$3d^{10}(^{1}_{0}S)\,4s^{2}\,4p^{6}\,4d^{8}(^{3}_{2}F)~^{3}F$ & 38488        & 4.7719   \\ 
9   & 0     & +   & 0.87~$3d^{10}(^{1}_{0}S)\,4s^{2}\,4p^{6}\,4d^{8}(^{1}_{0}S)~^{1}S$ + 0.08~$3d^{10}(^{1}_{0}S)\,4s^{2}\,4p^{6}\,4d^{8}(^{3}_{2}P)~^{3}P$ & 83632        & 10.3691  \\
\end{longtable}

\clearpage

\begin{longtable}{@{}@{\extracolsep{\fill}} lll p{0.65\linewidth} rr@{}}
\caption{I$^{10+}$ levels, 4d$^7$ valence electronic configuration (Tc-like ion).}\label{tab:ene-i10+}
No. & $J$ & $p$ & $LS$-composition &  $E$ (cm$^{-1}$) & $E$ (eV) \\ 
\midrule
\endfirsthead
\toprule
No. & $J$ & $p$ & $LS$-composition &  $E$ (cm$^{-1}$) & $E$ (eV) \\ 
\midrule
\endhead
\midrule
\endfoot
\bottomrule
\endlastfoot
1   & 9/2   & +   & 0.87~$3d^{10}(^{1}_{0}S)\,4s^{2}\,4p^{6}\,4d^{7}(^{4}_{3}F)~^{4}F$ + 0.09~$3d^{10}(^{1}_{0}S)\,4s^{2}\,4p^{6}\,4d^{7}(^{2}_{3}G)~^{2}G$ & 0            & 0.0000   \\ 
2   & 7/2   & +   & 0.94~$3d^{10}(^{1}_{0}S)\,4s^{2}\,4p^{6}\,4d^{7}(^{4}_{3}F)~^{4}F$ + 0.02~$3d^{10}(^{1}_{0}S)\,4s^{2}\,4p^{6}\,4d^{7}(^{2}_{3}G)~^{2}G$ & 11708        & 1.4516   \\ 
3   & 5/2   & +   & 0.79~$3d^{10}(^{1}_{0}S)\,4s^{2}\,4p^{6}\,4d^{7}(^{4}_{3}F)~^{4}F$ + 0.08~$3d^{10}(^{1}_{0}S)\,4s^{2}\,4p^{6}\,4d^{7}(^{2}_{3}D)~^{2}D$ + 0.05~$3d^{10}(^{1}_{0}S)\,4s^{2}\,4p^{6}\,4d^{7}(^{2}_{1}D)~^{2}D$ & 16052        & 1.9902   \\ 
4   & 3/2   & +   & 0.51~$3d^{10}(^{1}_{0}S)\,4s^{2}\,4p^{6}\,4d^{7}(^{4}_{3}F)~^{4}F$ + 0.19~$3d^{10}(^{1}_{0}S)\,4s^{2}\,4p^{6}\,4d^{7}(^{2}_{3}P)~^{2}P$ + 0.16~$3d^{10}(^{1}_{0}S)\,4s^{2}\,4p^{6}\,4d^{7}(^{2}_{3}D)~^{2}D$ & 17576        & 2.1791   \\ 
5   & 3/2   & +   & 0.41~$3d^{10}(^{1}_{0}S)\,4s^{2}\,4p^{6}\,4d^{7}(^{4}_{3}P)~^{4}P$ + 0.29~$3d^{10}(^{1}_{0}S)\,4s^{2}\,4p^{6}\,4d^{7}(^{4}_{3}F)~^{4}F$ + 0.26~$3d^{10}(^{1}_{0}S)\,4s^{2}\,4p^{6}\,4d^{7}(^{2}_{3}P)~^{2}P$ & 23999        & 2.9755   \\ 
6   & 5/2   & +   & 0.86~$3d^{10}(^{1}_{0}S)\,4s^{2}\,4p^{6}\,4d^{7}(^{4}_{3}P)~^{4}P$ + 0.06~$3d^{10}(^{1}_{0}S)\,4s^{2}\,4p^{6}\,4d^{7}(^{4}_{3}F)~^{4}F$ + 0.03~$3d^{10}(^{1}_{0}S)\,4s^{2}\,4p^{6}\,4d^{7}(^{2}_{1}D)~^{2}D$ & 27006        & 3.3483   \\ 
7   & 9/2   & +   & 0.64~$3d^{10}(^{1}_{0}S)\,4s^{2}\,4p^{6}\,4d^{7}(^{2}_{3}G)~^{2}G$ + 0.23~$3d^{10}(^{1}_{0}S)\,4s^{2}\,4p^{6}\,4d^{7}(^{2}_{3}H)~^{2}H$ + 0.09~$3d^{10}(^{1}_{0}S)\,4s^{2}\,4p^{6}\,4d^{7}(^{4}_{3}F)~^{4}F$ & 30024        & 3.7225   \\ 
8   & 1/2   & +   & 0.69~$3d^{10}(^{1}_{0}S)\,4s^{2}\,4p^{6}\,4d^{7}(^{4}_{3}P)~^{4}P$ + 0.27~$3d^{10}(^{1}_{0}S)\,4s^{2}\,4p^{6}\,4d^{7}(^{2}_{3}P)~^{2}P$ & 32539        & 4.0344   \\ 
9   & 7/2   & +   & 0.88~$3d^{10}(^{1}_{0}S)\,4s^{2}\,4p^{6}\,4d^{7}(^{2}_{3}G)~^{2}G$ + 0.07~$3d^{10}(^{1}_{0}S)\,4s^{2}\,4p^{6}\,4d^{7}(^{2}_{3}F)~^{2}F$ & 39893        & 4.9461   \\ 
10  & 11/2  & +   & 0.96~$3d^{10}(^{1}_{0}S)\,4s^{2}\,4p^{6}\,4d^{7}(^{2}_{3}H)~^{2}H$                         & 40353        & 5.0032   \\ 
11  & 3/2   & +   & 0.41~$3d^{10}(^{1}_{0}S)\,4s^{2}\,4p^{6}\,4d^{7}(^{4}_{3}P)~^{4}P$ + 0.24~$3d^{10}(^{1}_{0}S)\,4s^{2}\,4p^{6}\,4d^{7}(^{2}_{3}P)~^{2}P$ + 0.18~$3d^{10}(^{1}_{0}S)\,4s^{2}\,4p^{6}\,4d^{7}(^{2}_{3}D)~^{2}D$ & 42753        & 5.3007   \\ 
12  & 5/2   & +   & 0.61~$3d^{10}(^{1}_{0}S)\,4s^{2}\,4p^{6}\,4d^{7}(^{2}_{3}D)~^{2}D$ + 0.18~$3d^{10}(^{1}_{0}S)\,4s^{2}\,4p^{6}\,4d^{7}(^{2}_{1}D)~^{2}D$ + 0.10~$3d^{10}(^{1}_{0}S)\,4s^{2}\,4p^{6}\,4d^{7}(^{4}_{3}F)~^{4}F$ & 45583        & 5.6516   \\ 
13  & 1/2   & +   & 0.69~$3d^{10}(^{1}_{0}S)\,4s^{2}\,4p^{6}\,4d^{7}(^{2}_{3}P)~^{2}P$ + 0.27~$3d^{10}(^{1}_{0}S)\,4s^{2}\,4p^{6}\,4d^{7}(^{4}_{3}P)~^{4}P$ & 49927        & 6.1902   \\ 
14  & 9/2   & +   & 0.73~$3d^{10}(^{1}_{0}S)\,4s^{2}\,4p^{6}\,4d^{7}(^{2}_{3}H)~^{2}H$ + 0.23~$3d^{10}(^{1}_{0}S)\,4s^{2}\,4p^{6}\,4d^{7}(^{2}_{3}G)~^{2}G$ & 52753        & 6.5405   \\ 
15  & 5/2   & +   & 0.88~$3d^{10}(^{1}_{0}S)\,4s^{2}\,4p^{6}\,4d^{7}(^{2}_{3}F)~^{2}F$ + 0.05~$3d^{10}(^{1}_{0}S)\,4s^{2}\,4p^{6}\,4d^{7}(^{2}_{3}D)~^{2}D$ & 58169        & 7.2121   \\ 
16  & 3/2   & +   & 0.57~$3d^{10}(^{1}_{0}S)\,4s^{2}\,4p^{6}\,4d^{7}(^{2}_{3}D)~^{2}D$ + 0.27~$3d^{10}(^{1}_{0}S)\,4s^{2}\,4p^{6}\,4d^{7}(^{2}_{3}P)~^{2}P$ + 0.06~$3d^{10}(^{1}_{0}S)\,4s^{2}\,4p^{6}\,4d^{7}(^{4}_{3}P)~^{4}P$ & 62197        & 7.7114   \\ 
17  & 7/2   & +   & 0.89~$3d^{10}(^{1}_{0}S)\,4s^{2}\,4p^{6}\,4d^{7}(^{2}_{3}F)~^{2}F$ + 0.06~$3d^{10}(^{1}_{0}S)\,4s^{2}\,4p^{6}\,4d^{7}(^{2}_{3}G)~^{2}G$ & 65442        & 8.1137   \\ 
18  & 3/2   & +   & 0.89~$3d^{10}(^{1}_{0}S)\,4s^{2}\,4p^{6}\,4d^{7}(^{2}_{1}D)~^{2}D$ + 0.05~$3d^{10}(^{1}_{0}S)\,4s^{2}\,4p^{6}\,4d^{7}(^{2}_{3}D)~^{2}D$ & 92481        & 11.4662  \\ 
19  & 5/2   & +   & 0.68~$3d^{10}(^{1}_{0}S)\,4s^{2}\,4p^{6}\,4d^{7}(^{2}_{1}D)~^{2}D$ + 0.21~$3d^{10}(^{1}_{0}S)\,4s^{2}\,4p^{6}\,4d^{7}(^{2}_{3}D)~^{2}D$ + 0.06~$3d^{10}(^{1}_{0}S)\,4s^{2}\,4p^{6}\,4d^{7}(^{2}_{3}F)~^{2}F$ & 97525        & 12.0916  \\ 
\end{longtable}

\clearpage

\begin{longtable}{@{}@{\extracolsep{\fill}} lll p{0.65\linewidth} rr@{}}
\caption{I$^{11+}$ levels, 4d$^6$ valence electronic configuration (Mo-like ion).}\label{tab:ene-i11+}
No. & $J$ & $p$ & $LS$-composition &  $E$ (cm$^{-1}$) & $E$ (eV) \\ 
\midrule
\endfirsthead
\toprule
No. & $J$ & $p$ & $LS$-composition &  $E$ (cm$^{-1}$) & $E$ (eV) \\ 
\midrule
\endhead
\midrule
\endfoot
\bottomrule
\endlastfoot
1   & 4     & +   & 0.86~$3d^{10}(^{1}_{0}S)\,4s^{2}\,4p^{6}\,4d^{6}(^{5}_{4}D)~^{5}D$ + 0.06~$3d^{10}(^{1}_{0}S)\,4s^{2}\,4p^{6}\,4d^{6}(^{3}_{4}F)~^{3}F$ + 0.04~$3d^{10}(^{1}_{0}S)\,4s^{2}\,4p^{6}\,4d^{6}(^{3}_{2}F)~^{3}F$ & 0            & 0.0000   \\ 
2   & 3     & +   & 0.91~$3d^{10}(^{1}_{0}S)\,4s^{2}\,4p^{6}\,4d^{6}(^{5}_{4}D)~^{5}D$ + 0.02~$3d^{10}(^{1}_{0}S)\,4s^{2}\,4p^{6}\,4d^{6}(^{3}_{4}F)~^{3}F$ & 8593         & 1.0654   \\ 
3   & 2     & +   & 0.74~$3d^{10}(^{1}_{0}S)\,4s^{2}\,4p^{6}\,4d^{6}(^{5}_{4}D)~^{5}D$ + 0.09~$3d^{10}(^{1}_{0}S)\,4s^{2}\,4p^{6}\,4d^{6}(^{3}_{4}P)~^{3}P$ + 0.08~$3d^{10}(^{1}_{0}S)\,4s^{2}\,4p^{6}\,4d^{6}(^{3}_{2}P)~^{3}P$ & 9788         & 1.2136   \\ 
4   & 1     & +   & 0.85~$3d^{10}(^{1}_{0}S)\,4s^{2}\,4p^{6}\,4d^{6}(^{5}_{4}D)~^{5}D$ + 0.06~$3d^{10}(^{1}_{0}S)\,4s^{2}\,4p^{6}\,4d^{6}(^{3}_{2}P)~^{3}P$ + 0.04~$3d^{10}(^{1}_{0}S)\,4s^{2}\,4p^{6}\,4d^{6}(^{3}_{4}P)~^{3}P$ & 13920        & 1.7259   \\ 
5   & 0     & +   & 0.83~$3d^{10}(^{1}_{0}S)\,4s^{2}\,4p^{6}\,4d^{6}(^{5}_{4}D)~^{5}D$ + 0.08~$3d^{10}(^{1}_{0}S)\,4s^{2}\,4p^{6}\,4d^{6}(^{3}_{2}P)~^{3}P$ + 0.03~$3d^{10}(^{1}_{0}S)\,4s^{2}\,4p^{6}\,4d^{6}(^{3}_{4}P)~^{3}P$ & 14988        & 1.8582   \\ 
6   & 4     & +   & 0.26~$3d^{10}(^{1}_{0}S)\,4s^{2}\,4p^{6}\,4d^{6}(^{3}_{4}H)~^{3}H$ + 0.24~$3d^{10}(^{1}_{0}S)\,4s^{2}\,4p^{6}\,4d^{6}(^{3}_{4}G)~^{3}G$ + 0.15~$3d^{10}(^{1}_{0}S)\,4s^{2}\,4p^{6}\,4d^{6}(^{3}_{4}F)~^{3}F$ & 29831        & 3.6985   \\ 
7   & 2     & +   & 0.40~$3d^{10}(^{1}_{0}S)\,4s^{2}\,4p^{6}\,4d^{6}(^{3}_{4}P)~^{3}P$ + 0.21~$3d^{10}(^{1}_{0}S)\,4s^{2}\,4p^{6}\,4d^{6}(^{5}_{4}D)~^{5}D$ + 0.20~$3d^{10}(^{1}_{0}S)\,4s^{2}\,4p^{6}\,4d^{6}(^{3}_{2}P)~^{3}P$ & 34613        & 4.2915   \\ 
8   & 5     & +   & 0.54~$3d^{10}(^{1}_{0}S)\,4s^{2}\,4p^{6}\,4d^{6}(^{3}_{4}H)~^{3}H$ + 0.43~$3d^{10}(^{1}_{0}S)\,4s^{2}\,4p^{6}\,4d^{6}(^{3}_{4}G)~^{3}G$ & 35459        & 4.3964   \\ 
9   & 6     & +   & 0.85~$3d^{10}(^{1}_{0}S)\,4s^{2}\,4p^{6}\,4d^{6}(^{3}_{4}H)~^{3}H$ + 0.11~$3d^{10}(^{1}_{0}S)\,4s^{2}\,4p^{6}\,4d^{6}(^{1}_{4}I)~^{1}I$ & 35605        & 4.4144   \\ 
10  & 3     & +   & 0.44~$3d^{10}(^{1}_{0}S)\,4s^{2}\,4p^{6}\,4d^{6}(^{3}_{4}F)~^{3}F$ + 0.31~$3d^{10}(^{1}_{0}S)\,4s^{2}\,4p^{6}\,4d^{6}(^{3}_{4}G)~^{3}G$ + 0.14~$3d^{10}(^{1}_{0}S)\,4s^{2}\,4p^{6}\,4d^{6}(^{3}_{2}F)~^{3}F$ & 40865        & 5.0667   \\ 
11  & 2     & +   & 0.71~$3d^{10}(^{1}_{0}S)\,4s^{2}\,4p^{6}\,4d^{6}(^{3}_{4}F)~^{3}F$ + 0.11~$3d^{10}(^{1}_{0}S)\,4s^{2}\,4p^{6}\,4d^{6}(^{1}_{4}D)~^{1}D$ + 0.06~$3d^{10}(^{1}_{0}S)\,4s^{2}\,4p^{6}\,4d^{6}(^{3}_{2}F)~^{3}F$ & 42147        & 5.2255   \\ 
12  & 4     & +   & 0.54~$3d^{10}(^{1}_{0}S)\,4s^{2}\,4p^{6}\,4d^{6}(^{3}_{4}H)~^{3}H$ + 0.28~$3d^{10}(^{1}_{0}S)\,4s^{2}\,4p^{6}\,4d^{6}(^{3}_{4}F)~^{3}F$ + 0.10~$3d^{10}(^{1}_{0}S)\,4s^{2}\,4p^{6}\,4d^{6}(^{3}_{2}F)~^{3}F$ & 45975        & 5.7002   \\ 
13  & 0     & +   & 0.37~$3d^{10}(^{1}_{0}S)\,4s^{2}\,4p^{6}\,4d^{6}(^{1}_{4}S)~^{1}S$ + 0.22~$3d^{10}(^{1}_{0}S)\,4s^{2}\,4p^{6}\,4d^{6}(^{3}_{4}P)~^{3}P$ + 0.16~$3d^{10}(^{1}_{0}S)\,4s^{2}\,4p^{6}\,4d^{6}(^{3}_{2}P)~^{3}P$ & 50419        & 6.2512   \\ 
14  & 5     & +   & 0.53~$3d^{10}(^{1}_{0}S)\,4s^{2}\,4p^{6}\,4d^{6}(^{3}_{4}G)~^{3}G$ + 0.43~$3d^{10}(^{1}_{0}S)\,4s^{2}\,4p^{6}\,4d^{6}(^{3}_{4}H)~^{3}H$ & 51332        & 6.3644   \\ 
15  & 1     & +   & 0.52~$3d^{10}(^{1}_{0}S)\,4s^{2}\,4p^{6}\,4d^{6}(^{3}_{4}D)~^{3}D$ + 0.34~$3d^{10}(^{1}_{0}S)\,4s^{2}\,4p^{6}\,4d^{6}(^{3}_{4}P)~^{3}P$ + 0.08~$3d^{10}(^{1}_{0}S)\,4s^{2}\,4p^{6}\,4d^{6}(^{5}_{4}D)~^{5}D$ & 51518        & 6.3874   \\ 
16  & 4     & +   & 0.65~$3d^{10}(^{1}_{0}S)\,4s^{2}\,4p^{6}\,4d^{6}(^{3}_{4}G)~^{3}G$ + 0.13~$3d^{10}(^{1}_{0}S)\,4s^{2}\,4p^{6}\,4d^{6}(^{3}_{4}F)~^{3}F$ + 0.12~$3d^{10}(^{1}_{0}S)\,4s^{2}\,4p^{6}\,4d^{6}(^{1}_{2}G)~^{1}G$ & 52619        & 6.5239   \\ 
17  & 2     & +   & 0.70~$3d^{10}(^{1}_{0}S)\,4s^{2}\,4p^{6}\,4d^{6}(^{3}_{4}D)~^{3}D$ + 0.09~$3d^{10}(^{1}_{0}S)\,4s^{2}\,4p^{6}\,4d^{6}(^{3}_{2}P)~^{3}P$ + 0.06~$3d^{10}(^{1}_{0}S)\,4s^{2}\,4p^{6}\,4d^{6}(^{3}_{2}F)~^{3}F$ & 54384        & 6.7428   \\ 
18  & 3     & +   & 0.41~$3d^{10}(^{1}_{0}S)\,4s^{2}\,4p^{6}\,4d^{6}(^{3}_{4}G)~^{3}G$ + 0.37~$3d^{10}(^{1}_{0}S)\,4s^{2}\,4p^{6}\,4d^{6}(^{3}_{4}F)~^{3}F$ + 0.10~$3d^{10}(^{1}_{0}S)\,4s^{2}\,4p^{6}\,4d^{6}(^{3}_{4}D)~^{3}D$ & 55886        & 6.9289   \\ 
19  & 3     & +   & 0.73~$3d^{10}(^{1}_{0}S)\,4s^{2}\,4p^{6}\,4d^{6}(^{3}_{4}D)~^{3}D$ + 0.20~$3d^{10}(^{1}_{0}S)\,4s^{2}\,4p^{6}\,4d^{6}(^{3}_{4}G)~^{3}G$ + 0.02~$3d^{10}(^{1}_{0}S)\,4s^{2}\,4p^{6}\,4d^{6}(^{3}_{2}F)~^{3}F$ & 56957        & 7.0617   \\ 
20  & 6     & +   & 0.85~$3d^{10}(^{1}_{0}S)\,4s^{2}\,4p^{6}\,4d^{6}(^{1}_{4}I)~^{1}I$ + 0.11~$3d^{10}(^{1}_{0}S)\,4s^{2}\,4p^{6}\,4d^{6}(^{3}_{4}H)~^{3}H$ & 60399        & 7.4885   \\ 
21  & 1     & +   & 0.41~$3d^{10}(^{1}_{0}S)\,4s^{2}\,4p^{6}\,4d^{6}(^{3}_{4}D)~^{3}D$ + 0.35~$3d^{10}(^{1}_{0}S)\,4s^{2}\,4p^{6}\,4d^{6}(^{3}_{4}P)~^{3}P$ + 0.17~$3d^{10}(^{1}_{0}S)\,4s^{2}\,4p^{6}\,4d^{6}(^{3}_{2}P)~^{3}P$ & 60946        & 7.5563   \\ 
22  & 0     & +   & 0.67~$3d^{10}(^{1}_{0}S)\,4s^{2}\,4p^{6}\,4d^{6}(^{3}_{4}P)~^{3}P$ + 0.19~$3d^{10}(^{1}_{0}S)\,4s^{2}\,4p^{6}\,4d^{6}(^{3}_{2}P)~^{3}P$ + 0.05~$3d^{10}(^{1}_{0}S)\,4s^{2}\,4p^{6}\,4d^{6}(^{1}_{0}S)~^{1}S$ & 67808        & 8.4072   \\ 
23  & 4     & +   & 0.59~$3d^{10}(^{1}_{0}S)\,4s^{2}\,4p^{6}\,4d^{6}(^{1}_{4}G)~^{1}G$ + 0.16~$3d^{10}(^{1}_{0}S)\,4s^{2}\,4p^{6}\,4d^{6}(^{3}_{4}F)~^{3}F$ + 0.12~$3d^{10}(^{1}_{0}S)\,4s^{2}\,4p^{6}\,4d^{6}(^{3}_{4}H)~^{3}H$ & 68988        & 8.5534   \\ 
24  & 3     & +   & 0.66~$3d^{10}(^{1}_{0}S)\,4s^{2}\,4p^{6}\,4d^{6}(^{1}_{4}F)~^{1}F$ + 0.23~$3d^{10}(^{1}_{0}S)\,4s^{2}\,4p^{6}\,4d^{6}(^{3}_{2}F)~^{3}F$ + 0.05~$3d^{10}(^{1}_{0}S)\,4s^{2}\,4p^{6}\,4d^{6}(^{3}_{4}F)~^{3}F$ & 70632        & 8.7573   \\ 
25  & 2     & +   & 0.54~$3d^{10}(^{1}_{0}S)\,4s^{2}\,4p^{6}\,4d^{6}(^{1}_{4}D)~^{1}D$ + 0.12~$3d^{10}(^{1}_{0}S)\,4s^{2}\,4p^{6}\,4d^{6}(^{3}_{4}F)~^{3}F$ + 0.09~$3d^{10}(^{1}_{0}S)\,4s^{2}\,4p^{6}\,4d^{6}(^{1}_{2}D)~^{1}D$ & 71747        & 8.8955   \\ 
26  & 1     & +   & 0.70~$3d^{10}(^{1}_{0}S)\,4s^{2}\,4p^{6}\,4d^{6}(^{3}_{2}P)~^{3}P$ + 0.22~$3d^{10}(^{1}_{0}S)\,4s^{2}\,4p^{6}\,4d^{6}(^{3}_{4}P)~^{3}P$ & 82643        & 10.2464  \\ 
27  & 4     & +   & 0.70~$3d^{10}(^{1}_{0}S)\,4s^{2}\,4p^{6}\,4d^{6}(^{3}_{2}F)~^{3}F$ + 0.13~$3d^{10}(^{1}_{0}S)\,4s^{2}\,4p^{6}\,4d^{6}(^{3}_{4}F)~^{3}F$ + 0.08~$3d^{10}(^{1}_{0}S)\,4s^{2}\,4p^{6}\,4d^{6}(^{1}_{2}G)~^{1}G$ & 88256        & 10.9424  \\ 
28  & 2     & +   & 0.81~$3d^{10}(^{1}_{0}S)\,4s^{2}\,4p^{6}\,4d^{6}(^{3}_{2}F)~^{3}F$ + 0.09~$3d^{10}(^{1}_{0}S)\,4s^{2}\,4p^{6}\,4d^{6}(^{3}_{4}F)~^{3}F$ + 0.04~$3d^{10}(^{1}_{0}S)\,4s^{2}\,4p^{6}\,4d^{6}(^{3}_{4}D)~^{3}D$ & 89378        & 11.0815  \\ 
29  & 0     & +   & 0.51~$3d^{10}(^{1}_{0}S)\,4s^{2}\,4p^{6}\,4d^{6}(^{3}_{2}P)~^{3}P$ + 0.37~$3d^{10}(^{1}_{0}S)\,4s^{2}\,4p^{6}\,4d^{6}(^{1}_{4}S)~^{1}S$ + 0.03~$3d^{10}(^{1}_{0}S)\,4s^{2}\,4p^{6}\,4d^{6}(^{3}_{4}P)~^{3}P$ & 90368        & 11.2042  \\ 
30  & 2     & +   & 0.52~$3d^{10}(^{1}_{0}S)\,4s^{2}\,4p^{6}\,4d^{6}(^{3}_{2}P)~^{3}P$ + 0.29~$3d^{10}(^{1}_{0}S)\,4s^{2}\,4p^{6}\,4d^{6}(^{3}_{4}P)~^{3}P$ + 0.06~$3d^{10}(^{1}_{0}S)\,4s^{2}\,4p^{6}\,4d^{6}(^{1}_{4}D)~^{1}D$ & 97150        & 12.0451  \\ 
31  & 3     & +   & 0.56~$3d^{10}(^{1}_{0}S)\,4s^{2}\,4p^{6}\,4d^{6}(^{3}_{2}F)~^{3}F$ + 0.22~$3d^{10}(^{1}_{0}S)\,4s^{2}\,4p^{6}\,4d^{6}(^{1}_{4}F)~^{1}F$ + 0.08~$3d^{10}(^{1}_{0}S)\,4s^{2}\,4p^{6}\,4d^{6}(^{3}_{4}F)~^{3}F$ & 97262        & 12.0590  \\ 
32  & 4     & +   & 0.61~$3d^{10}(^{1}_{0}S)\,4s^{2}\,4p^{6}\,4d^{6}(^{1}_{2}G)~^{1}G$ + 0.22~$3d^{10}(^{1}_{0}S)\,4s^{2}\,4p^{6}\,4d^{6}(^{1}_{4}G)~^{1}G$ + 0.06~$3d^{10}(^{1}_{0}S)\,4s^{2}\,4p^{6}\,4d^{6}(^{3}_{4}F)~^{3}F$ & 102037       & 12.6510  \\ 
33  & 2     & +   & 0.76~$3d^{10}(^{1}_{0}S)\,4s^{2}\,4p^{6}\,4d^{6}(^{1}_{2}D)~^{1}D$ + 0.18~$3d^{10}(^{1}_{0}S)\,4s^{2}\,4p^{6}\,4d^{6}(^{1}_{4}D)~^{1}D$ & 127612       & 15.8219  \\ 
34  & 0     & +   & 0.77~$3d^{10}(^{1}_{0}S)\,4s^{2}\,4p^{6}\,4d^{6}(^{1}_{0}S)~^{1}S$ + 0.16~$3d^{10}(^{1}_{0}S)\,4s^{2}\,4p^{6}\,4d^{6}(^{1}_{4}S)~^{1}S$ + 0.02~$3d^{10}(^{1}_{0}S)\,4s^{2}\,4p^{6}\,4d^{6}(^{3}_{2}P)~^{3}P$ & 164497       & 20.3950  \\ 
\end{longtable}

\clearpage

\begin{longtable}{@{}@{\extracolsep{\fill}} lll p{0.65\linewidth} rr@{}}
\caption{I$^{12+}$ levels, 4d$^5$ valence electronic configuration (Nb-like ion).}\label{tab:ene-i12+}
No. & $J$ & $p$ & $LS$-composition &  $E$ (cm$^{-1}$) & $E$ (eV) \\ 
\midrule
\endfirsthead
\toprule
No. & $J$ & $p$ & $LS$-composition &  $E$ (cm$^{-1}$) & $E$ (eV) \\ 
\midrule
\endhead
\midrule
\endfoot
\bottomrule
\endlastfoot
1   & 5/2   & +   & 0.88~$3d^{10}(^{1}_{0}S)\,4s^{2}\,4p^{6}\,4d^{5}(^{6}_{5}S)~^{6}S$ + 0.08~$3d^{10}(^{1}_{0}S)\,4s^{2}\,4p^{6}\,4d^{5}(^{4}_{3}P)~^{4}P$ & 0            & 0.0000   \\ 
2   & 5/2   & +   & 0.34~$3d^{10}(^{1}_{0}S)\,4s^{2}\,4p^{6}\,4d^{5}(^{4}_{5}G)~^{4}G$ + 0.15~$3d^{10}(^{1}_{0}S)\,4s^{2}\,4p^{6}\,4d^{5}(^{2}_{3}F)~^{2}F$ + 0.15~$3d^{10}(^{1}_{0}S)\,4s^{2}\,4p^{6}\,4d^{5}(^{4}_{5}D)~^{4}D$ & 37613        & 4.6634   \\ 
3   & 7/2   & +   & 0.72~$3d^{10}(^{1}_{0}S)\,4s^{2}\,4p^{6}\,4d^{5}(^{4}_{5}G)~^{4}G$ + 0.09~$3d^{10}(^{1}_{0}S)\,4s^{2}\,4p^{6}\,4d^{5}(^{4}_{5}D)~^{4}D$ + 0.09~$3d^{10}(^{1}_{0}S)\,4s^{2}\,4p^{6}\,4d^{5}(^{2}_{3}F)~^{2}F$ & 44372        & 5.5014   \\ 
4   & 11/2  & +   & 0.84~$3d^{10}(^{1}_{0}S)\,4s^{2}\,4p^{6}\,4d^{5}(^{4}_{5}G)~^{4}G$ + 0.10~$3d^{10}(^{1}_{0}S)\,4s^{2}\,4p^{6}\,4d^{5}(^{2}_{3}H)~^{2}H$ + 0.02~$3d^{10}(^{1}_{0}S)\,4s^{2}\,4p^{6}\,4d^{5}(^{2}_{5}I)~^{2}I$ & 46119        & 5.7180   \\ 
5   & 5/2   & +   & 0.43~$3d^{10}(^{1}_{0}S)\,4s^{2}\,4p^{6}\,4d^{5}(^{4}_{5}G)~^{4}G$ + 0.23~$3d^{10}(^{1}_{0}S)\,4s^{2}\,4p^{6}\,4d^{5}(^{4}_{5}D)~^{4}D$ + 0.21~$3d^{10}(^{1}_{0}S)\,4s^{2}\,4p^{6}\,4d^{5}(^{4}_{3}P)~^{4}P$ & 46333        & 5.7445   \\ 
6   & 3/2   & +   & 0.45~$3d^{10}(^{1}_{0}S)\,4s^{2}\,4p^{6}\,4d^{5}(^{4}_{3}P)~^{4}P$ + 0.44~$3d^{10}(^{1}_{0}S)\,4s^{2}\,4p^{6}\,4d^{5}(^{4}_{5}D)~^{4}D$ + 0.03~$3d^{10}(^{1}_{0}S)\,4s^{2}\,4p^{6}\,4d^{5}(^{4}_{3}F)~^{4}F$ & 46461        & 5.7605   \\ 
7   & 9/2   & +   & 0.87~$3d^{10}(^{1}_{0}S)\,4s^{2}\,4p^{6}\,4d^{5}(^{4}_{5}G)~^{4}G$ + 0.05~$3d^{10}(^{1}_{0}S)\,4s^{2}\,4p^{6}\,4d^{5}(^{4}_{3}F)~^{4}F$ + 0.03~$3d^{10}(^{1}_{0}S)\,4s^{2}\,4p^{6}\,4d^{5}(^{2}_{3}H)~^{2}H$ & 47255        & 5.8589   \\ 
8   & 1/2   & +   & 0.61~$3d^{10}(^{1}_{0}S)\,4s^{2}\,4p^{6}\,4d^{5}(^{4}_{3}P)~^{4}P$ + 0.32~$3d^{10}(^{1}_{0}S)\,4s^{2}\,4p^{6}\,4d^{5}(^{4}_{5}D)~^{4}D$ + 0.04~$3d^{10}(^{1}_{0}S)\,4s^{2}\,4p^{6}\,4d^{5}(^{2}_{5}S)~^{2}S$ & 51380        & 6.3703   \\ 
9   & 7/2   & +   & 0.73~$3d^{10}(^{1}_{0}S)\,4s^{2}\,4p^{6}\,4d^{5}(^{4}_{5}D)~^{4}D$ + 0.16~$3d^{10}(^{1}_{0}S)\,4s^{2}\,4p^{6}\,4d^{5}(^{4}_{5}G)~^{4}G$ + 0.06~$3d^{10}(^{1}_{0}S)\,4s^{2}\,4p^{6}\,4d^{5}(^{4}_{3}F)~^{4}F$ & 55605        & 6.8941   \\ 
10  & 5/2   & +   & 0.40~$3d^{10}(^{1}_{0}S)\,4s^{2}\,4p^{6}\,4d^{5}(^{4}_{5}D)~^{4}D$ + 0.19~$3d^{10}(^{1}_{0}S)\,4s^{2}\,4p^{6}\,4d^{5}(^{4}_{3}P)~^{4}P$ + 0.17~$3d^{10}(^{1}_{0}S)\,4s^{2}\,4p^{6}\,4d^{5}(^{2}_{5}D)~^{2}D$ & 60412        & 7.4902   \\ 
11  & 1/2   & +   & 0.64~$3d^{10}(^{1}_{0}S)\,4s^{2}\,4p^{6}\,4d^{5}(^{4}_{5}D)~^{4}D$ + 0.30~$3d^{10}(^{1}_{0}S)\,4s^{2}\,4p^{6}\,4d^{5}(^{4}_{3}P)~^{4}P$ + 0.02~$3d^{10}(^{1}_{0}S)\,4s^{2}\,4p^{6}\,4d^{5}(^{2}_{5}S)~^{2}S$ & 62506        & 7.7497   \\ 
12  & 3/2   & +   & 0.44~$3d^{10}(^{1}_{0}S)\,4s^{2}\,4p^{6}\,4d^{5}(^{4}_{5}D)~^{4}D$ + 0.44~$3d^{10}(^{1}_{0}S)\,4s^{2}\,4p^{6}\,4d^{5}(^{4}_{3}P)~^{4}P$ + 0.04~$3d^{10}(^{1}_{0}S)\,4s^{2}\,4p^{6}\,4d^{5}(^{2}_{5}D)~^{2}D$ & 64755        & 8.0286   \\ 
13  & 11/2  & +   & 0.75~$3d^{10}(^{1}_{0}S)\,4s^{2}\,4p^{6}\,4d^{5}(^{2}_{5}I)~^{2}I$ + 0.14~$3d^{10}(^{1}_{0}S)\,4s^{2}\,4p^{6}\,4d^{5}(^{2}_{3}H)~^{2}H$ + 0.08~$3d^{10}(^{1}_{0}S)\,4s^{2}\,4p^{6}\,4d^{5}(^{4}_{5}G)~^{4}G$ & 66233        & 8.2118   \\ 
14  & 7/2   & +   & 0.33~$3d^{10}(^{1}_{0}S)\,4s^{2}\,4p^{6}\,4d^{5}(^{4}_{3}F)~^{4}F$ + 0.22~$3d^{10}(^{1}_{0}S)\,4s^{2}\,4p^{6}\,4d^{5}(^{2}_{3}F)~^{2}F$ + 0.20~$3d^{10}(^{1}_{0}S)\,4s^{2}\,4p^{6}\,4d^{5}(^{2}_{5}F)~^{2}F$ & 70003        & 8.6793   \\ 
15  & 13/2  & +   & 0.97~$3d^{10}(^{1}_{0}S)\,4s^{2}\,4p^{6}\,4d^{5}(^{2}_{5}I)~^{2}I$                         & 71684        & 8.8876   \\ 
16  & 3/2   & +   & 0.47~$3d^{10}(^{1}_{0}S)\,4s^{2}\,4p^{6}\,4d^{5}(^{4}_{3}F)~^{4}F$ + 0.30~$3d^{10}(^{1}_{0}S)\,4s^{2}\,4p^{6}\,4d^{5}(^{2}_{5}D)~^{2}D$ + 0.13~$3d^{10}(^{1}_{0}S)\,4s^{2}\,4p^{6}\,4d^{5}(^{2}_{1}D)~^{2}D$ & 72030        & 8.9306   \\ 
17  & 9/2   & +   & 0.43~$3d^{10}(^{1}_{0}S)\,4s^{2}\,4p^{6}\,4d^{5}(^{2}_{5}G)~^{2}G$ + 0.37~$3d^{10}(^{1}_{0}S)\,4s^{2}\,4p^{6}\,4d^{5}(^{4}_{3}F)~^{4}F$ + 0.08~$3d^{10}(^{1}_{0}S)\,4s^{2}\,4p^{6}\,4d^{5}(^{2}_{3}H)~^{2}H$ & 72135        & 8.9435   \\ 
18  & 5/2   & +   & 0.20~$3d^{10}(^{1}_{0}S)\,4s^{2}\,4p^{6}\,4d^{5}(^{4}_{3}F)~^{4}F$ + 0.19~$3d^{10}(^{1}_{0}S)\,4s^{2}\,4p^{6}\,4d^{5}(^{4}_{3}P)~^{4}P$ + 0.13~$3d^{10}(^{1}_{0}S)\,4s^{2}\,4p^{6}\,4d^{5}(^{2}_{3}F)~^{2}F$ & 73357        & 9.0951   \\ 
19  & 5/2   & +   & 0.41~$3d^{10}(^{1}_{0}S)\,4s^{2}\,4p^{6}\,4d^{5}(^{4}_{3}F)~^{4}F$ + 0.22~$3d^{10}(^{1}_{0}S)\,4s^{2}\,4p^{6}\,4d^{5}(^{2}_{3}F)~^{2}F$ + 0.11~$3d^{10}(^{1}_{0}S)\,4s^{2}\,4p^{6}\,4d^{5}(^{4}_{5}D)~^{4}D$ & 79823        & 9.8968   \\ 
20  & 9/2   & +   & 0.50~$3d^{10}(^{1}_{0}S)\,4s^{2}\,4p^{6}\,4d^{5}(^{2}_{3}H)~^{2}H$ + 0.39~$3d^{10}(^{1}_{0}S)\,4s^{2}\,4p^{6}\,4d^{5}(^{4}_{3}F)~^{4}F$ + 0.07~$3d^{10}(^{1}_{0}S)\,4s^{2}\,4p^{6}\,4d^{5}(^{2}_{5}G)~^{2}G$ & 83044        & 10.2962  \\ 
21  & 7/2   & +   & 0.49~$3d^{10}(^{1}_{0}S)\,4s^{2}\,4p^{6}\,4d^{5}(^{2}_{5}G)~^{2}G$ + 0.18~$3d^{10}(^{1}_{0}S)\,4s^{2}\,4p^{6}\,4d^{5}(^{2}_{5}F)~^{2}F$ + 0.15~$3d^{10}(^{1}_{0}S)\,4s^{2}\,4p^{6}\,4d^{5}(^{4}_{3}F)~^{4}F$ & 84872        & 10.5228  \\ 
22  & 3/2   & +   & 0.43~$3d^{10}(^{1}_{0}S)\,4s^{2}\,4p^{6}\,4d^{5}(^{2}_{5}D)~^{2}D$ + 0.42~$3d^{10}(^{1}_{0}S)\,4s^{2}\,4p^{6}\,4d^{5}(^{4}_{3}F)~^{4}F$ + 0.04~$3d^{10}(^{1}_{0}S)\,4s^{2}\,4p^{6}\,4d^{5}(^{2}_{1}D)~^{2}D$ & 87237        & 10.8160  \\ 
23  & 5/2   & +   & 0.57~$3d^{10}(^{1}_{0}S)\,4s^{2}\,4p^{6}\,4d^{5}(^{2}_{5}F)~^{2}F$ + 0.18~$3d^{10}(^{1}_{0}S)\,4s^{2}\,4p^{6}\,4d^{5}(^{4}_{3}F)~^{4}F$ + 0.11~$3d^{10}(^{1}_{0}S)\,4s^{2}\,4p^{6}\,4d^{5}(^{2}_{3}D)~^{2}D$ & 89208        & 11.0604  \\ 
24  & 7/2   & +   & 0.53~$3d^{10}(^{1}_{0}S)\,4s^{2}\,4p^{6}\,4d^{5}(^{2}_{3}F)~^{2}F$ + 0.23~$3d^{10}(^{1}_{0}S)\,4s^{2}\,4p^{6}\,4d^{5}(^{2}_{5}G)~^{2}G$ + 0.07~$3d^{10}(^{1}_{0}S)\,4s^{2}\,4p^{6}\,4d^{5}(^{4}_{3}F)~^{4}F$ & 89254        & 11.0661  \\ 
25  & 7/2   & +   & 0.50~$3d^{10}(^{1}_{0}S)\,4s^{2}\,4p^{6}\,4d^{5}(^{2}_{5}F)~^{2}F$ + 0.30~$3d^{10}(^{1}_{0}S)\,4s^{2}\,4p^{6}\,4d^{5}(^{4}_{3}F)~^{4}F$ + 0.08~$3d^{10}(^{1}_{0}S)\,4s^{2}\,4p^{6}\,4d^{5}(^{2}_{3}G)~^{2}G$ & 94894        & 11.7653  \\ 
26  & 11/2  & +   & 0.73~$3d^{10}(^{1}_{0}S)\,4s^{2}\,4p^{6}\,4d^{5}(^{2}_{3}H)~^{2}H$ + 0.19~$3d^{10}(^{1}_{0}S)\,4s^{2}\,4p^{6}\,4d^{5}(^{2}_{5}I)~^{2}I$ + 0.05~$3d^{10}(^{1}_{0}S)\,4s^{2}\,4p^{6}\,4d^{5}(^{4}_{5}G)~^{4}G$ & 95688        & 11.8638  \\ 
27  & 9/2   & +   & 0.42~$3d^{10}(^{1}_{0}S)\,4s^{2}\,4p^{6}\,4d^{5}(^{2}_{5}G)~^{2}G$ + 0.35~$3d^{10}(^{1}_{0}S)\,4s^{2}\,4p^{6}\,4d^{5}(^{2}_{3}H)~^{2}H$ + 0.15~$3d^{10}(^{1}_{0}S)\,4s^{2}\,4p^{6}\,4d^{5}(^{4}_{3}F)~^{4}F$ & 96867        & 12.0100  \\ 
28  & 1/2   & +   & 0.85~$3d^{10}(^{1}_{0}S)\,4s^{2}\,4p^{6}\,4d^{5}(^{2}_{5}S)~^{2}S$ + 0.06~$3d^{10}(^{1}_{0}S)\,4s^{2}\,4p^{6}\,4d^{5}(^{4}_{3}P)~^{4}P$ + 0.05~$3d^{10}(^{1}_{0}S)\,4s^{2}\,4p^{6}\,4d^{5}(^{2}_{3}P)~^{2}P$ & 97669        & 12.1095  \\ 
29  & 5/2   & +   & 0.37~$3d^{10}(^{1}_{0}S)\,4s^{2}\,4p^{6}\,4d^{5}(^{2}_{3}F)~^{2}F$ + 0.31~$3d^{10}(^{1}_{0}S)\,4s^{2}\,4p^{6}\,4d^{5}(^{2}_{5}D)~^{2}D$ + 0.09~$3d^{10}(^{1}_{0}S)\,4s^{2}\,4p^{6}\,4d^{5}(^{2}_{3}D)~^{2}D$ & 102865       & 12.7537  \\ 
30  & 3/2   & +   & 0.86~$3d^{10}(^{1}_{0}S)\,4s^{2}\,4p^{6}\,4d^{5}(^{2}_{3}D)~^{2}D$ + 0.04~$3d^{10}(^{1}_{0}S)\,4s^{2}\,4p^{6}\,4d^{5}(^{2}_{1}D)~^{2}D$ + 0.04~$3d^{10}(^{1}_{0}S)\,4s^{2}\,4p^{6}\,4d^{5}(^{4}_{5}D)~^{4}D$ & 109658       & 13.5958  \\ 
31  & 5/2   & +   & 0.69~$3d^{10}(^{1}_{0}S)\,4s^{2}\,4p^{6}\,4d^{5}(^{2}_{3}D)~^{2}D$ + 0.15~$3d^{10}(^{1}_{0}S)\,4s^{2}\,4p^{6}\,4d^{5}(^{2}_{5}F)~^{2}F$ + 0.05~$3d^{10}(^{1}_{0}S)\,4s^{2}\,4p^{6}\,4d^{5}(^{4}_{5}D)~^{4}D$ & 115818       & 14.3596  \\ 
32  & 9/2   & +   & 0.92~$3d^{10}(^{1}_{0}S)\,4s^{2}\,4p^{6}\,4d^{5}(^{2}_{3}G)~^{2}G$ + 0.03~$3d^{10}(^{1}_{0}S)\,4s^{2}\,4p^{6}\,4d^{5}(^{2}_{5}G)~^{2}G$ & 121622       & 15.0792  \\ 
33  & 7/2   & +   & 0.84~$3d^{10}(^{1}_{0}S)\,4s^{2}\,4p^{6}\,4d^{5}(^{2}_{3}G)~^{2}G$ + 0.07~$3d^{10}(^{1}_{0}S)\,4s^{2}\,4p^{6}\,4d^{5}(^{2}_{5}F)~^{2}F$ + 0.04~$3d^{10}(^{1}_{0}S)\,4s^{2}\,4p^{6}\,4d^{5}(^{2}_{5}G)~^{2}G$ & 124367       & 15.4196  \\ 
34  & 3/2   & +   & 0.84~$3d^{10}(^{1}_{0}S)\,4s^{2}\,4p^{6}\,4d^{5}(^{2}_{3}P)~^{2}P$ + 0.08~$3d^{10}(^{1}_{0}S)\,4s^{2}\,4p^{6}\,4d^{5}(^{2}_{1}D)~^{2}D$ + 0.03~$3d^{10}(^{1}_{0}S)\,4s^{2}\,4p^{6}\,4d^{5}(^{2}_{5}D)~^{2}D$ & 140213       & 17.3842  \\ 
35  & 1/2   & +   & 0.90~$3d^{10}(^{1}_{0}S)\,4s^{2}\,4p^{6}\,4d^{5}(^{2}_{3}P)~^{2}P$ + 0.05~$3d^{10}(^{1}_{0}S)\,4s^{2}\,4p^{6}\,4d^{5}(^{2}_{5}S)~^{2}S$ & 145774       & 18.0737  \\ 
36  & 5/2   & +   & 0.76~$3d^{10}(^{1}_{0}S)\,4s^{2}\,4p^{6}\,4d^{5}(^{2}_{1}D)~^{2}D$ + 0.16~$3d^{10}(^{1}_{0}S)\,4s^{2}\,4p^{6}\,4d^{5}(^{2}_{5}D)~^{2}D$ + 0.02~$3d^{10}(^{1}_{0}S)\,4s^{2}\,4p^{6}\,4d^{5}(^{2}_{3}D)~^{2}D$ & 158534       & 19.6558  \\ 
37  & 3/2   & +   & 0.65~$3d^{10}(^{1}_{0}S)\,4s^{2}\,4p^{6}\,4d^{5}(^{2}_{1}D)~^{2}D$ + 0.15~$3d^{10}(^{1}_{0}S)\,4s^{2}\,4p^{6}\,4d^{5}(^{2}_{5}D)~^{2}D$ + 0.12~$3d^{10}(^{1}_{0}S)\,4s^{2}\,4p^{6}\,4d^{5}(^{2}_{3}P)~^{2}P$ & 162282       & 20.1204  \\ 
\end{longtable}

\clearpage

\begin{longtable}{@{}@{\extracolsep{\fill}} ll SS @{}}
\caption{M1 transition for I$^{8+}$ ion.}\label{tab:tr-i8+}
Upper level No. & Upper level No. &  {$\lambda$ (nm)} & {$A$ (s$^{-1}$)} \\ 
\midrule
\endfirsthead
\toprule
Upper level No. & Upper level No. &  {$\lambda$ (nm)} & {$A$ (s$^{-1}$)} \\
\midrule
\endhead
\midrule
\endfoot
\bottomrule
\endlastfoot
2 & 1 & 695.29 & 48.09 \\
\end{longtable}

\clearpage

\begin{longtable}{@{}@{\extracolsep{\fill}} ll SS @{}}
\caption{Selected M1 transitions for I$^{9+}$ ion.}\label{tab:tr-i9+}
Upper level No. & Upper level No. &  {$\lambda$ (nm)} & {$A$ (s$^{-1}$)} \\
\midrule
\endfirsthead
\toprule
Upper level No. & Upper level No. &  {$\lambda$ (nm)} & {$A$ (s$^{-1}$)} \\
\midrule
\endhead
\midrule
\endfoot
\bottomrule
\endlastfoot
9 & 6 & 193.56 & 6.37E+02 \\
7 & 1 & 260.21 & 4.67E+01 \\
8 & 3 & 393.14 & 1.05E+02 \\
6 & 2 & 513.82 & 3.02E+01 \\
8 & 4 & 691.66 & 3.88E+01 \\
3 & 1 & 766.17 & 5.63E+01 \\
4 & 2 & 867.74 & 2.17E+01 \\
4 & 3 & 910.90 & 2.01E+01 \\
\end{longtable}

\clearpage

\begin{longtable}{@{}@{\extracolsep{\fill}} ll SS @{}}
\caption{Selected M1 transitions for I$^{10+}$ ion.}\label{tab:tr-i10+}
Upper level No. & Upper level No. &  {$\lambda$ (nm)} & {$A$ (s$^{-1}$)} \\
\midrule
\endfirsthead
\toprule
Upper level No. & Upper level No. &  {$\lambda$ (nm)} & {$A$ (s$^{-1}$)} \\
\midrule
\endhead
\midrule
\endfoot
\bottomrule
\endlastfoot
19 & 2 & 116.53 & 1.84E+01 \\
18 & 3 & 130.84 & 1.01E+01 \\
18 & 4 & 133.50 & 4.48E+01 \\
19 & 5 & 136.01 & 1.39E+01 \\
19 & 6 & 141.81 & 1.62E+02 \\
18 & 6 & 152.73 & 6.12E+01 \\
17 & 1 & 152.81 & 1.17E+01 \\
19 & 9 & 173.51 & 1.40E+01 \\
19 & 11 & 182.58 & 4.16E+01 \\
17 & 2 & 186.10 & 2.35E+01 \\
19 & 12 & 192.52 & 1.25E+01 \\
18 & 11 & 201.09 & 8.69E+01 \\
18 & 12 & 213.23 & 1.77E+02 \\
16 & 3 & 216.71 & 8.46E+01 \\
18 & 13 & 234.99 & 1.24E+01 \\
15 & 3 & 237.43 & 1.35E+01 \\
15 & 4 & 246.34 & 1.55E+01 \\
9 & 1 & 250.67 & 1.01E+01 \\
19 & 15 & 254.09 & 9.58E+01 \\
16 & 5 & 261.80 & 1.19E+01 \\
17 & 7 & 282.34 & 4.24E+01 \\
19 & 16 & 283.06 & 2.76E+01 \\
18 & 15 & 291.44 & 2.68E+01 \\
12 & 2 & 295.20 & 2.23E+02 \\
13 & 4 & 309.11 & 1.10E+01 \\
19 & 17 & 311.69 & 2.69E+01 \\
7 & 1 & 333.07 & 8.97E+01 \\
9 & 2 & 354.80 & 2.44E+01 \\
11 & 3 & 374.52 & 2.85E+01 \\
17 & 9 & 391.41 & 2.63E+01 \\
9 & 3 & 419.45 & 1.12E+01 \\
14 & 7 & 439.96 & 6.68E+01 \\
12 & 5 & 463.31 & 2.89E+01 \\
16 & 11 & 514.31 & 7.72E+01 \\
11 & 5 & 533.22 & 7.33E+01 \\
12 & 6 & 538.30 & 1.84E+01 \\
13 & 8 & 575.12 & 8.59E+01 \\
16 & 12 & 601.91 & 5.75E+01 \\
11 & 6 & 635.04 & 6.23E+01 \\
6 & 2 & 653.68 & 1.34E+01 \\
8 & 4 & 668.28 & 2.28E+01 \\
14 & 10 & 806.48 & 2.12E+01 \\
2 & 1 & 854.11 & 5.06E+01 \\
9 & 7 & 1013.23 & 1.02E+01 \\
8 & 5 & 1170.93 & 1.48E+01 \\
5 & 3 & 1258.37 & 1.08E+01 \\
\end{longtable}

\clearpage

\begin{longtable}{@{}@{\extracolsep{\fill}} ll SS @{}}
\caption{Selected M1 transitions for I$^{11+}$ ion.}\label{tab:tr-i11+}
Upper level No. & Upper level No. &  {$\lambda$ (nm)} & {$A$ (s$^{-1}$)} \\
\midrule
\endfirsthead
\toprule
Upper level No. & Upper level No. &  {$\lambda$ (nm)} & {$A$ (s$^{-1}$)} \\
\midrule
\endhead
\midrule
\endfoot
\bottomrule
\endlastfoot
34 & 21 & 96.57 & 9.75E+01 \\
33 & 7 & 107.53 & 1.30E+01 \\
31 & 2 & 112.78 & 2.18E+01 \\
30 & 2 & 112.92 & 8.37E+01 \\
27 & 1 & 113.31 & 2.65E+01 \\
31 & 3 & 114.32 & 1.02E+01 \\
33 & 10 & 115.28 & 1.44E+01 \\
34 & 26 & 122.17 & 7.66E+02 \\
27 & 2 & 125.53 & 2.07E+01 \\
28 & 3 & 125.64 & 1.07E+01 \\
29 & 4 & 130.81 & 1.18E+02 \\
33 & 15 & 131.42 & 8.75E+01 \\
26 & 3 & 137.26 & 1.16E+02 \\
33 & 18 & 139.42 & 2.15E+01 \\
33 & 19 & 141.53 & 8.57E+01 \\
26 & 5 & 147.81 & 2.42E+01 \\
33 & 21 & 150.00 & 1.87E+01 \\
25 & 2 & 158.34 & 2.31E+01 \\
30 & 7 & 159.91 & 1.11E+01 \\
27 & 6 & 171.16 & 2.71E+01 \\
33 & 24 & 175.50 & 2.08E+01 \\
19 & 1 & 175.57 & 7.12E+01 \\
31 & 10 & 177.31 & 1.83E+01 \\
32 & 12 & 178.37 & 9.65E+01 \\
31 & 11 & 181.44 & 1.31E+01 \\
28 & 7 & 182.60 & 1.96E+01 \\
22 & 4 & 185.57 & 7.32E+01 \\
27 & 8 & 189.41 & 1.52E+01 \\
16 & 1 & 190.05 & 1.81E+01 \\
31 & 12 & 194.98 & 1.90E+01 \\
21 & 3 & 195.47 & 7.58E+01 \\
32 & 14 & 197.22 & 8.88E+01 \\
32 & 16 & 202.36 & 3.24E+01 \\
28 & 10 & 206.13 & 4.14E+01 \\
19 & 2 & 206.77 & 1.97E+01 \\
26 & 7 & 208.20 & 1.74E+02 \\
18 & 2 & 211.45 & 5.03E+01 \\
21 & 4 & 212.65 & 4.17E+01 \\
32 & 18 & 216.68 & 9.09E+01 \\
12 & 1 & 217.51 & 6.89E+01 \\
31 & 16 & 224.00 & 2.30E+01 \\
17 & 3 & 224.24 & 4.13E+01 \\
27 & 12 & 236.51 & 5.64E+01 \\
15 & 3 & 239.64 & 1.04E+01 \\
30 & 18 & 242.34 & 3.29E+01 \\
10 & 1 & 244.71 & 3.44E+01 \\
24 & 6 & 245.09 & 2.70E+01 \\
31 & 19 & 248.10 & 1.72E+02 \\
30 & 19 & 248.80 & 5.79E+01 \\
29 & 15 & 257.40 & 3.14E+02 \\
33 & 28 & 261.55 & 1.20E+01 \\
28 & 15 & 264.13 & 3.28E+01 \\
15 & 4 & 265.97 & 2.69E+01 \\
25 & 7 & 269.30 & 1.75E+01 \\
27 & 14 & 270.83 & 7.15E+01 \\
15 & 5 & 273.75 & 7.43E+01 \\
13 & 4 & 273.98 & 2.83E+02 \\
30 & 21 & 276.21 & 2.17E+01 \\
28 & 17 & 285.76 & 7.46E+01 \\
11 & 2 & 298.03 & 2.19E+01 \\
23 & 8 & 298.25 & 6.39E+01 \\
11 & 3 & 309.04 & 4.50E+01 \\
10 & 2 & 309.87 & 4.95E+01 \\
26 & 13 & 310.33 & 2.24E+01 \\
32 & 24 & 318.42 & 2.10E+01 \\
27 & 19 & 319.49 & 3.15E+01 \\
26 & 15 & 321.28 & 1.80E+01 \\
10 & 3 & 321.78 & 1.19E+01 \\
25 & 10 & 323.82 & 6.06E+01 \\
6 & 1 & 335.23 & 9.02E+01 \\
25 & 11 & 337.83 & 6.97E+01 \\
29 & 21 & 339.88 & 2.99E+02 \\
28 & 21 & 351.71 & 1.02E+01 \\
23 & 10 & 355.59 & 3.10E+01 \\
19 & 6 & 368.65 & 2.99E+01 \\
21 & 7 & 379.76 & 1.06E+02 \\
7 & 2 & 384.32 & 2.05E+02 \\
30 & 25 & 393.65 & 6.98E+01 \\
20 & 8 & 400.97 & 2.30E+01 \\
20 & 9 & 403.32 & 5.02E+01 \\
23 & 12 & 434.54 & 9.09E+01 \\
19 & 7 & 447.56 & 2.06E+01 \\
18 & 7 & 470.10 & 1.28E+01 \\
7 & 4 & 483.25 & 4.37E+01 \\
25 & 15 & 494.33 & 1.90E+01 \\
17 & 7 & 505.79 & 1.19E+01 \\
28 & 24 & 533.45 & 6.16E+01 \\
23 & 14 & 566.39 & 1.38E+01 \\
27 & 24 & 567.41 & 2.44E+01 \\
16 & 8 & 582.76 & 2.41E+01 \\
15 & 7 & 591.56 & 1.07E+02 \\
22 & 15 & 613.85 & 4.05E+01 \\
12 & 6 & 619.41 & 7.37E+01 \\
19 & 10 & 621.46 & 2.14E+01 \\
14 & 8 & 630.00 & 2.22E+01 \\
25 & 18 & 630.46 & 1.37E+01 \\
14 & 9 & 635.83 & 4.05E+01 \\
18 & 10 & 665.77 & 1.44E+01 \\
26 & 22 & 674.10 & 3.62E+01 \\
30 & 26 & 689.32 & 3.77E+01 \\
18 & 11 & 727.86 & 2.11E+01 \\
17 & 10 & 739.70 & 1.01E+01 \\
10 & 6 & 906.23 & 1.85E+01 \\
21 & 15 & 1060.66 & 1.03E+01 \\
31 & 27 & 1110.37 & 1.23E+01 \\
2 & 1 & 1163.71 & 2.07E+01 \\
\end{longtable}

\clearpage

\begin{longtable}{@{}@{\extracolsep{\fill}} ll SS @{}}
\caption{Selected M1 transitions for I$^{12+}$ ion.}\label{tab:tr-i12+}
Upper level No. & Upper level No. &  {$\lambda$ (nm)} & {$A$ (s$^{-1}$)} \\
\midrule
\endfirsthead
\toprule
Upper level No. & Upper level No. &  {$\lambda$ (nm)} & {$A$ (s$^{-1}$)} \\
\midrule
\endhead
\midrule
\endfoot
\bottomrule
\endlastfoot
36 & 3 & 87.59 & 1.22E+01 \\
36 & 5 & 89.13 & 2.23E+01 \\
36 & 6 & 89.23 & 1.64E+01 \\
37 & 8 & 90.17 & 1.31E+01 \\
37 & 10 & 98.16 & 4.02E+01 \\
35 & 6 & 100.69 & 2.31E+01 \\
36 & 10 & 101.91 & 3.71E+01 \\
37 & 12 & 102.54 & 3.43E+01 \\
34 & 5 & 106.52 & 1.94E+01 \\
37 & 16 & 110.80 & 5.28E+01 \\
36 & 14 & 112.95 & 2.37E+01 \\
33 & 2 & 115.27 & 1.01E+01 \\
36 & 18 & 117.40 & 1.06E+02 \\
35 & 11 & 120.09 & 2.61E+01 \\
37 & 19 & 121.27 & 7.27E+01 \\
35 & 12 & 123.43 & 7.02E+01 \\
19 & 1 & 125.28 & 3.02E+01 \\
34 & 10 & 125.31 & 2.28E+01 \\
36 & 19 & 127.05 & 4.14E+01 \\
33 & 5 & 128.15 & 1.39E+01 \\
32 & 3 & 129.45 & 1.26E+01 \\
32 & 4 & 132.45 & 3.76E+01 \\
37 & 22 & 133.25 & 2.11E+01 \\
36 & 21 & 135.75 & 4.41E+01 \\
18 & 1 & 136.32 & 4.45E+01 \\
37 & 23 & 136.85 & 1.62E+02 \\
31 & 3 & 139.96 & 2.36E+01 \\
31 & 5 & 143.91 & 1.80E+01 \\
31 & 6 & 144.18 & 4.60E+01 \\
36 & 23 & 144.25 & 2.31E+01 \\
36 & 24 & 144.34 & 7.89E+01 \\
34 & 16 & 146.66 & 2.81E+01 \\
29 & 2 & 153.25 & 2.82E+01 \\
12 & 1 & 154.43 & 1.30E+02 \\
37 & 28 & 154.77 & 1.96E+01 \\
36 & 25 & 157.13 & 4.47E+01 \\
30 & 5 & 157.92 & 1.71E+01 \\
30 & 6 & 158.24 & 2.12E+01 \\
10 & 1 & 165.53 & 1.23E+02 \\
34 & 19 & 165.59 & 2.38E+01 \\
31 & 9 & 166.08 & 1.26E+02 \\
29 & 3 & 170.96 & 2.03E+01 \\
30 & 8 & 171.59 & 5.82E+01 \\
36 & 29 & 179.63 & 1.07E+01 \\
31 & 10 & 180.49 & 1.09E+01 \\
33 & 17 & 191.45 & 4.51E+01 \\
32 & 14 & 193.73 & 5.34E+01 \\
28 & 6 & 195.28 & 1.63E+02 \\
31 & 12 & 195.84 & 2.68E+01 \\
33 & 18 & 196.04 & 1.42E+01 \\
34 & 23 & 196.06 & 1.46E+01 \\
26 & 4 & 201.74 & 8.38E+01 \\
36 & 30 & 204.60 & 1.71E+02 \\
25 & 5 & 205.93 & 1.37E+01 \\
26 & 7 & 206.47 & 4.71E+01 \\
35 & 28 & 207.88 & 2.13E+02 \\
29 & 9 & 211.59 & 8.13E+01 \\
30 & 11 & 212.08 & 1.58E+02 \\
37 & 31 & 215.22 & 1.47E+02 \\
6 & 1 & 215.23 & 1.70E+02 \\
5 & 1 & 215.83 & 1.10E+02 \\
28 & 8 & 216.03 & 4.38E+01 \\
30 & 12 & 222.70 & 8.78E+01 \\
24 & 3 & 222.80 & 1.39E+01 \\
23 & 3 & 223.03 & 4.80E+01 \\
31 & 16 & 228.37 & 1.64E+01 \\
23 & 5 & 233.23 & 2.42E+01 \\
23 & 6 & 233.94 & 1.43E+01 \\
34 & 28 & 235.05 & 3.14E+01 \\
24 & 7 & 238.10 & 9.56E+01 \\
33 & 20 & 242.00 & 2.56E+01 \\
27 & 9 & 242.35 & 4.27E+01 \\
33 & 21 & 253.20 & 1.68E+01 \\
20 & 3 & 258.58 & 4.64E+01 \\
21 & 7 & 265.84 & 4.64E+01 \\
2 & 1 & 265.87 & 1.01E+02 \\
20 & 4 & 270.82 & 3.49E+01 \\
32 & 21 & 272.11 & 2.43E+01 \\
30 & 18 & 275.48 & 2.06E+01 \\
31 & 19 & 277.81 & 4.91E+01 \\
20 & 7 & 279.42 & 1.02E+02 \\
18 & 2 & 279.76 & 1.12E+01 \\
19 & 3 & 282.08 & 4.59E+01 \\
33 & 23 & 284.42 & 3.77E+01 \\
16 & 2 & 290.55 & 7.38E+01 \\
24 & 9 & 297.18 & 9.78E+01 \\
23 & 9 & 297.59 & 3.72E+01 \\
19 & 5 & 298.60 & 3.26E+01 \\
28 & 12 & 303.82 & 4.70E+01 \\
27 & 13 & 326.43 & 3.30E+01 \\
30 & 19 & 335.18 & 3.51E+01 \\
29 & 18 & 338.89 & 1.77E+02 \\
33 & 25 & 339.29 & 4.82E+01 \\
26 & 13 & 339.50 & 1.47E+02 \\
21 & 9 & 341.68 & 2.36E+01 \\
18 & 3 & 345.00 & 3.42E+01 \\
24 & 10 & 346.72 & 2.47E+01 \\
23 & 10 & 347.27 & 1.02E+01 \\
31 & 22 & 349.88 & 1.07E+01 \\
17 & 3 & 360.19 & 1.64E+01 \\
33 & 27 & 363.64 & 3.42E+01 \\
20 & 9 & 364.44 & 1.23E+01 \\
12 & 2 & 368.43 & 1.17E+01 \\
18 & 5 & 370.03 & 2.59E+02 \\
18 & 6 & 371.80 & 1.39E+01 \\
22 & 10 & 372.79 & 3.61E+01 \\
32 & 25 & 374.14 & 3.08E+01 \\
31 & 23 & 375.80 & 5.90E+01 \\
31 & 24 & 376.45 & 2.50E+01 \\
17 & 4 & 384.38 & 2.41E+01 \\
16 & 6 & 391.10 & 8.60E+01 \\
25 & 14 & 401.76 & 2.70E+01 \\
22 & 11 & 404.34 & 2.43E+01 \\
23 & 12 & 408.95 & 1.05E+01 \\
19 & 9 & 412.91 & 1.08E+01 \\
26 & 15 & 416.59 & 3.96E+01 \\
26 & 17 & 424.57 & 1.97E+01 \\
29 & 19 & 433.98 & 5.60E+01 \\
10 & 2 & 438.61 & 1.66E+01 \\
25 & 17 & 439.38 & 3.68E+01 \\
14 & 7 & 439.60 & 2.43E+01 \\
22 & 12 & 444.80 & 4.43E+01 \\
37 & 34 & 453.12 & 3.46E+01 \\
31 & 25 & 477.91 & 2.66E+01 \\
16 & 8 & 484.25 & 1.56E+01 \\
30 & 23 & 489.01 & 1.05E+01 \\
13 & 4 & 497.17 & 3.92E+01 \\
19 & 10 & 515.19 & 5.33E+01 \\
13 & 7 & 526.94 & 1.01E+01 \\
12 & 6 & 546.64 & 3.97E+01 \\
29 & 21 & 555.77 & 1.29E+01 \\
18 & 9 & 563.30 & 1.50E+01 \\
23 & 16 & 582.15 & 3.44E+01 \\
24 & 17 & 584.13 & 3.61E+01 \\
17 & 9 & 604.96 & 1.23E+01 \\
11 & 6 & 623.28 & 2.48E+01 \\
10 & 3 & 623.42 & 2.10E+01 \\
24 & 18 & 629.06 & 2.04E+01 \\
23 & 18 & 630.88 & 1.11E+01 \\
29 & 22 & 639.87 & 2.81E+01 \\
22 & 16 & 657.60 & 1.62E+01 \\
25 & 19 & 663.53 & 1.51E+01 \\
21 & 14 & 672.54 & 1.74E+01 \\
27 & 20 & 723.42 & 3.61E+01 \\
29 & 24 & 734.69 & 1.89E+01 \\
27 & 21 & 833.66 & 1.11E+01 \\
9 & 3 & 890.24 & 1.17E+01 \\
11 & 8 & 898.82 & 4.24E+01 \\
20 & 17 & 916.63 & 1.47E+01 \\
19 & 14 & 1018.37 & 2.05E+01 \\
9 & 5 & 1078.52 & 1.22E+01 \\
6 & 2 & 1130.13 & 1.01E+01 \\
5 & 2 & 1146.81 & 2.41E+01 \\
\end{longtable}

\clearpage

\begin{figure*}[!t]
\centering
\includegraphics[width=\linewidth]{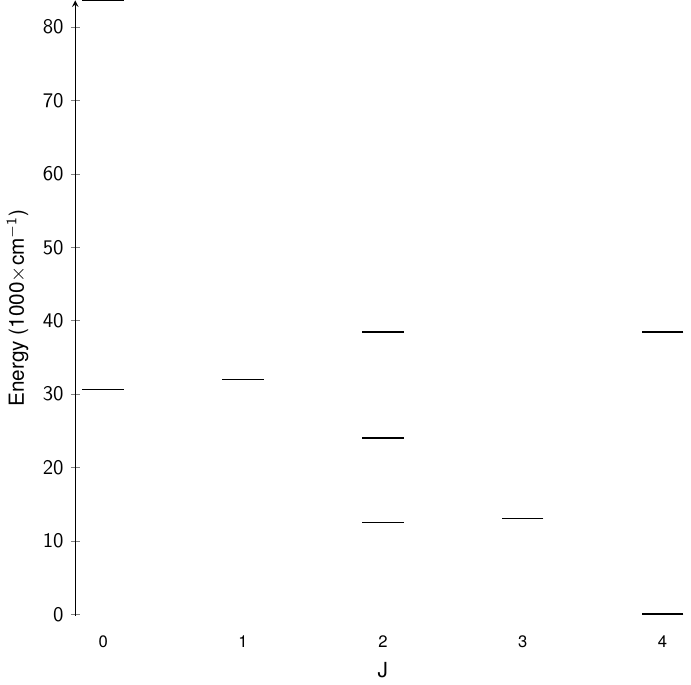}
\caption{Energy level diagram for all states of \ce{[Kr] {4d}^8} configuration of \ce{I^9+}. }
\label{fig:i9+_levels}
\end{figure*}

\clearpage

\begin{figure*}[!t]
\centering
\includegraphics[width=\linewidth]{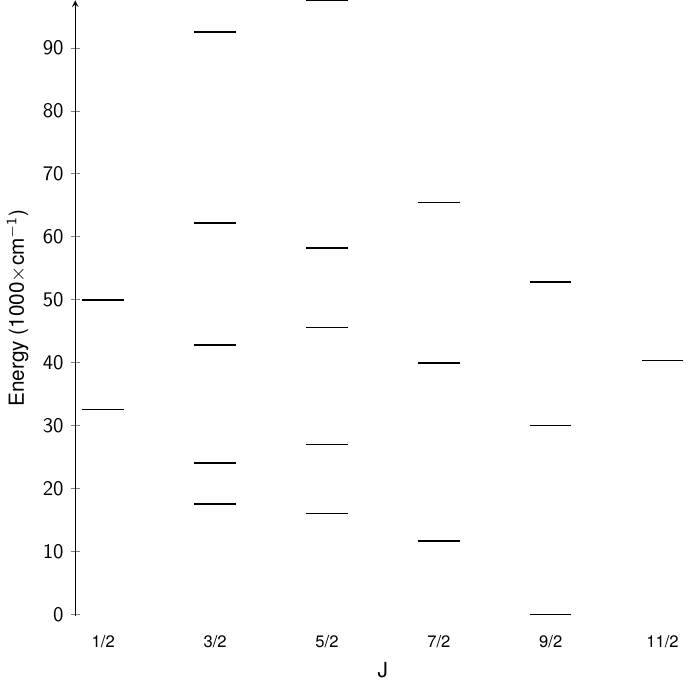}
\caption{Energy level diagram for all states of \ce{[Kr] {4d}^7} configuration of \ce{I^10+}. }
\label{fig:i10+_levels}
\end{figure*}

\clearpage

\begin{figure*}[!t]
\centering
\includegraphics[width=\linewidth]{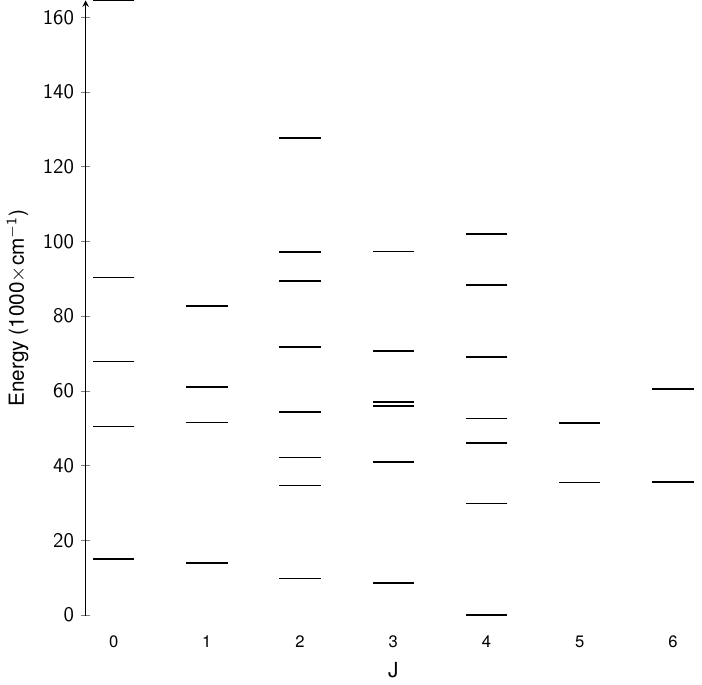}
\caption{Energy level diagram for all states of \ce{[Kr] {4d}^6} configuration of \ce{I^11+}. }
\label{fig:i11+_levels}
\end{figure*}

\clearpage

\begin{figure*}[!t]
\centering
\includegraphics[width=\linewidth]{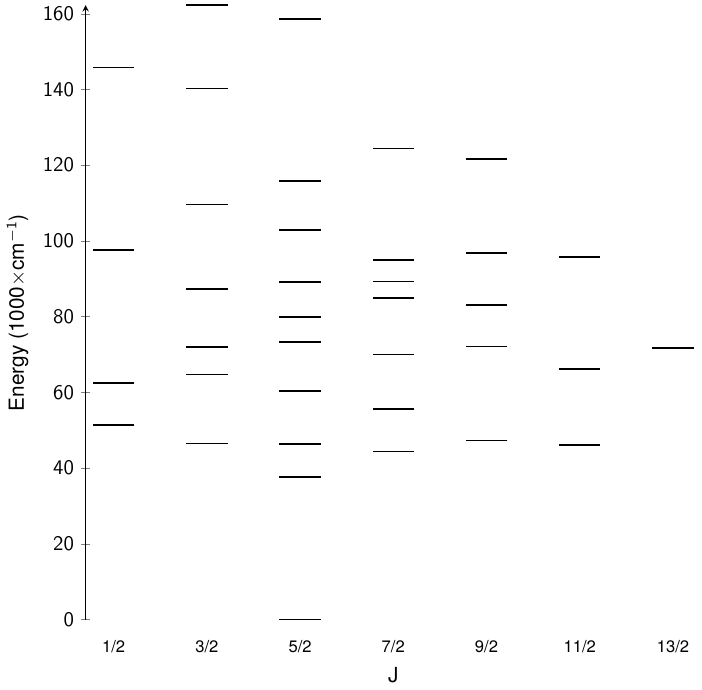}
\caption{Energy level diagram for all states of \ce{[Kr] {4d}^5} configuration of \ce{I^12+}. }
\label{fig:i12+_levels}
\end{figure*}

\end{document}